\documentclass[final,3p,authoryear,times]{elsarticle} %

\usepackage{amssymb}
\usepackage{amsthm}
\usepackage[latin1]{inputenc} 
\usepackage{subfigure} 
\usepackage{amsmath}
\usepackage{amsfonts}
\usepackage{latexsym}
\usepackage{textcomp}
\usepackage{graphicx}
\usepackage{tipa}

\usepackage{cancel} 
\usepackage{ctable}
\usepackage{appendix}
\usepackage{url}

\newtheorem{prop}{Proposition}
\newdefinition{rmk}{Remark}
\newproof{pf}{Proof}
\newproof{pot}{Proof of Theorem \ref{thm2}}
\newcommand{\real}{\text{\rm I\hspace{-0.6mm}R}}   

\def\bx{\boldsymbol{x}}

\def\bX{\boldsymbol{X}}

\def\b0{\boldsymbol{0}}
\def\b1{\boldsymbol{1}}

\def\bmu{\mbox{\boldmath $\mu$}}

\def\bpsi{\mbox{\boldmath $\psi$}}

\def\bbeta{\mbox{\boldmath $\beta$}}

\def\bxi{\mbox{\boldmath $\xi$}}

\def\bvartheta{\mbox{\boldmath $\vartheta$}}

\def\bvartheta{\mbox{\boldmath $\vartheta$}}

\def\bSigma{\mbox{\boldmath $\Sigma$}}

\journal{Computational Statistics and Data Analysis, \textbf{71}(4): 159--182, 2014}

\begin{document}

\begin{frontmatter}

\title{
Model-based clustering via\\ linear cluster-weighted models
}


\author[CT]{S.~Ingrassia\corref{cor}} 
\ead{s.ingrassia@unict.it}
\author[MI]{S.C.~Minotti}  
\ead{simona.minotti@unimib.it}
\author[CT]{A.~Punzo} 
\ead{antonio.punzo@unict.it}
\cortext[cor]{Corresponding author}
\address[CT]{Dipartimento di Economia e Impresa, Università di Catania, \\ Corso Italia 55, 95129 Catania (Italy)}
\address[MI]{Dipartimento di Statistica, Università di Milano-Bicocca (Italy)}

\address{}

\begin{abstract}
A novel family of twelve mixture models with random covariates, nested in the linear $t$ cluster-weighted model (CWM), is introduced for model-based clustering.
The linear $t$ CWM was recently presented as a robust alternative to the better known linear Gaussian CWM.
The proposed family of models provides a unified framework that also includes the linear Gaussian CWM as a special case.
Maximum likelihood parameter estimation is carried out within the EM framework, and both the BIC and the ICL are used for model selection.
A simple and effective hierarchical random initialization is also proposed for the EM algorithm. 
The novel model-based clustering technique is illustrated in some applications to real data. 
Finally, a simulation study for evaluating the performance of the BIC and the ICL is presented.
\end{abstract}


\end{frontmatter}


\section{Introduction}
\label{sec:introduction}

In direct applications of finite mixture models \citep[see][pp.~2--3]{Titt:Smit:Mako:stat:1985}, we assume that each mixture-component represents a group (or cluster) in the original data.
The term ``model-based clustering'' has been used to describe the adoption of mixture models for clustering or, more often, to describe the use of a family of mixture models for clustering (see \citealp{Fral:Raft:Howm:1998} and \citealp{McLa:Basf:mixt:1988}).
An overview of mixture models is given in \citet{EvHa:fini:1981}, \citet{Titt:Smit:Mako:stat:1985}, \citet{McLa:Peel:fini:2000}, and \citet{Fruh:Fine:2006}.

This paper focuses on data arising from a real-valued random vector $\left(Y,\bX'\right)':\Omega\rightarrow\real^{d+1}$, having joint density $p\left(y,\bx\right)$, where $Y$ is the response variable and $\bX$ is the vector of covariates.
Standard model-based clustering techniques assume that $\Omega$ can be partitioned into $G$ groups $\Omega_1,\ldots,\Omega_G$.
As for finite mixtures of linear regressions (see, e.g., \citealp{Leis:Flex:2004} and \citealp[][Chapter~8]{Fruh:Fine:2006}) we assume that, for each $\Omega_g$, the dependence of $Y$ on $\bx$ can be modeled by 
\begin{equation*}
Y=\mu\left(\bx;\bbeta_g\right)+\varepsilon_g=\beta_{0g}+\bbeta_{1g}'\bx+\varepsilon_g, 
\label{eq:regression model}
\end{equation*}
where $\bbeta_g=\left(\beta_{0g},\bbeta_{1g}'\right)'$, $\mu\left(\bx;\bbeta_g\right)=E\left(Y|\bX=\bx,\Omega_g\right)$ is the linear regression function and $\varepsilon_g$ 
is the error variable, independent with respect to $\bX$, with zero mean and finite constant variance $\sigma_g^2$, $g=1,\ldots,G$.
However, as highlighted in \citet{Henn:Iden:2000}, finite mixtures of linear regressions are inadequate for most of the applications because they assume \textit{assignment independence}: the probability for a point $\left(y,\bx'\right)'$ to be generated by one of the mixture components has to be the same for all covariates values $\bx$.
In other words, the assignment of the data points to the clusters has to be independent of the covariates.

Here, differently from finite mixtures of linear regressions, we assume random covariates having a parametric specification.
This allows for \textit{assignment dependence}: the covariate distributions of the mixture components can also be distinct. 
In the framework of mixture models with random covariates, the cluster weighted model \citep[CWM;][]{Gers:Nonl:1997}, with equation 
\begin{equation}
p\left(y,\bx\right)=\sum_{g=1}^G\pi_g  p\left(y,\bx |\Omega_g\right)=
\sum_{g=1}^G\pi_g p\left(y|\bx,\Omega_g\right)p\left(\bx|\Omega_g\right),
\label{eq:general cwm}
\end{equation}
also called saturated mixture regression model by \citet{Wede:Conc:2002}, constitutes a reference approach to model the joint density.
In \eqref{eq:general cwm}, normality of both $p\left(y|\bx,\Omega_g\right)$ and $p\left(\bx|\Omega_g\right)$ is commonly assumed (see, e.g., \citealp{Gers:Nonl:1997} and \citealp{Punz:Flex:2013}).
Alternatively, \citet{Ingr:Mino:Vitt:Loca:2012} propose also the use of the $t$ distribution which provides, as other approaches \citep{Punz:McNi:Robu:2013,Punz:McNi:Robu:2014,Punz:McNi:Robu:2014b}, more robust fitting for groups of observations with longer than normal tails or noise data (see, e.g., \citealp{Zell:Baye:1976}, \citealp{Lang:Litt:Tayl:Robu:1989}, \citealp{Peel:McLa:2000}, \citealp[][Chapter~7]{McLa:Peel:fini:2000}, \citealp{Chat:Varv:Robu:2008}, and \citealp{Gres:Ingr:Cons:2010}).
In particular, the authors consider
\begin{equation}
p\left(y|\bx,\Omega_g\right) = h_t\left(y|\bx; \bxi_g,\zeta_g\right)=\frac{\Gamma\left(\displaystyle\frac{\zeta_g+1}{2}\right)}{\left(\pi \zeta_g\sigma^2_g\right)^{\frac{1}{2}}\left\{1+\delta\left[y,\mu\left(\bx;\bbeta_g\right);\sigma^2_g\right]\right\}^{\frac{\zeta_g+1}{2}}}
\label{eq:tCWM y|x}
\end{equation}
and
\begin{equation}
p\left(\bx|\Omega_g\right) = h_{t_d}\left(\bx;\boldsymbol{\vartheta}_g,\nu_g\right)=\frac{\Gamma\left(\displaystyle\frac{\nu_g+d}{2}\right) \left| \bSigma_g\right|^{-\frac{1}{2}}}{\left(\pi \nu_g\right)^{\frac{d}{2}}\left[1+\delta\left(\bx,\bmu_g;\bSigma_g\right)\right]^{\frac{\nu_g+d}{2}}},
\label{eq:tCWM x}
\end{equation}
with $\bxi_g=\left\{\bbeta_g,\sigma^2_g\right\}$, $\boldsymbol{\vartheta}_g=\left\{\bmu_g,\bSigma_g\right\}$, $\delta\left[y,\mu(\bx;\bbeta_g);\sigma^2_g\right]=\left[y-\mu\left(\bx;\bbeta_g\right)\right]^2\big/\sigma^2_g$, and  
$\delta\left(\bx,\bmu_g;\bSigma_g\right)=\left(\bx-\bmu_g\right)'\bSigma_g^{-1}\left(\bx-\bmu_g\right)$.
Thus, \eqref{eq:tCWM y|x} is the density of a (generalized) univariate $t$ distribution, with location parameter $\mu\left(\bx ;\bbeta_g\right)$, scale parameter $\sigma^2_g$, and $\zeta_g$ degrees of freedom, while \eqref{eq:tCWM x} is the density of a multivariate $t$ distribution with location parameter $\bmu_g$, inner product matrix $\bSigma_g$, and $\nu_g$ degrees of freedom.
By substituting \eqref{eq:tCWM y|x} and \eqref{eq:tCWM x} into \eqref{eq:general cwm}, we obtain the linear $t$ CWM
\begin{equation}
p\left(y,\bx;\text{\textsubtilde{$\boldsymbol{\psi}$}}\right) = \sum_{g=1}^{G} \pi_g  h_t\left(y|\bx;\bxi_g,\zeta_g\right)h_{t_d}\left(\bx;\bvartheta_g,\nu_g\right),
\label{eq:linear $t$ CWM}
\end{equation}
where the set of all unknown parameters is denoted by $\text{\textsubtilde{$\boldsymbol{\psi}$}}=\left\{\bpsi_1,\ldots,\bpsi_G\right\}$,
with $\bpsi_g=\left\{\pi_g, \bxi_g,\zeta_g,\boldsymbol{\vartheta}_g,\nu_g\right\}$. 
Recent developments in CWMs can be found in \citet{Punz:Flex:2013}, \citet{Punz:McNi:Robu:2014}, \citet{Punz:Ingr:Clus:2015,Punz:Ingr:Pars:2015}, \citet{Sube:Punz:Ingr:McNi:Clus:2013,Sube:Punz:Ingr:McNi:Clus:2015}, and \citet{Ingr:Punz:Vitt:TheG:2015}.

In this paper, we introduce a family of twelve linear CWMs obtained from \eqref{eq:linear $t$ CWM} by imposing convenient component distributional constraints.
If $\zeta_g,\nu_g\rightarrow\infty$, the linear Gaussian (normal) CWM is obtained as a special case.
The resulting models are easily interpretable and appropriate for describing various practical situations.
In particular, they also allow us to infer if the group-structure of the data is due to the contribution of $\bX$, $Y|\bX$, or both. 

The paper is organized as follows. 
In Section~\ref{sec:Clustering via CWM}, we recall model-based clustering according to the CW approach, and give some preliminary results.
In Section~\ref{sec:The parsimonious Student-t LCWMs}, we introduce the novel family of models.
Model fitting in the EM paradigm is presented in Section~\ref{sec:Estimation via the EM algorithm}, related computational aspects are addressed in Section~\ref{sec:Computational issues and partitions evaluation}, and model selection is discussed in Section~\ref{sec:model selection}.   
In Section~\ref{sec:real datas} some applications to real data are illustrated. 
In Section~\ref{sec:Comparing the BIC and the ICL} simulations for a comparison between BIC and ICL are  described. 
Finally, in Section~\ref{sec:conclusions}, we give a summary of the paper and some directions for further research.

\section{Preliminary results for model-based clustering}
\label{sec:Clustering via CWM}

This section recalls some basic ideas on model-based clustering according to the CWM approach and provides some preliminary results that will be useful for definition and justification of our family of models. 

Let $\left(y_1,\bx_1'\right)',\ldots,\left(y_N,\bx_N'\right)'$ be a sample of size $N$ from \eqref{eq:linear $t$ CWM}.
Once \textsubtilde{$\bpsi$} is estimated (fixed), the posterior probability that the generic unit $\left(y_n,\bx_n'\right)'$, $n=1,\ldots,N$, comes from component $\Omega_g$ is given by 
\begin{equation}
\tau_{ng}=P\left(\Omega_g|y_n,\bx_n;\textsubtilde{$\bpsi$}\right)=\frac{\pi_g  h_t\left(y_n|\bx_n;\bxi_g,\zeta_g\right)h_{t_d}\left(\bx_n;\bvartheta_g,\nu_g\right)}{p\left(y_n,\bx_n;\textsubtilde{$\bpsi$}\right)}, \quad g=1,\ldots,G.
\label{eq:posterior probabilities}
\end{equation}
These probabilities, which depend on both marginal and conditional densities, represent the basis for clustering and classification.  

The following two propositions, which generalize some results given in \citet{Ingr:Mino:Vitt:Loca:2012}, require the preliminary definition of
\begin{equation}
p\left(y|\bx;\text{\textsubtilde{$\pi$}},\text{\textsubtilde{$\boldsymbol{\xi}$}},\text{\textsubtilde{$\zeta$}}\right) = \sum_{g=1}^{G} \pi_g  h_t\left(y|\bx;\bxi_g,\zeta_g\right)
\label{eq:linear $t$ FMR}
\end{equation}
and
\begin{equation}
p\left(\bx;\text{\textsubtilde{$\pi$}},\text{\textsubtilde{$\boldsymbol{\vartheta}$}},\text{\textsubtilde{$\nu$}}\right) = \sum_{g=1}^{G} \pi_g  h_{t_d}\left(\bx;\bvartheta_g,\nu_g\right),
\label{eq:$t$ FMD}
\end{equation}
which correspond to a finite mixture of linear $t$ regressions and a finite mixture of multivariate $t$ distributions ($\text{\textsubtilde{$\pi$}}=\left\{\pi_1,\ldots,\pi_{G-1}\right\}$, $\text{\textsubtilde{$\boldsymbol{\xi}$}}=\left\{\xi_1,\ldots,\xi_G\right\}$, $\text{\textsubtilde{$\zeta$}}=\left\{\zeta_1,\ldots,\zeta_G\right\}$, $\text{\textsubtilde{$\boldsymbol{\vartheta}$}}=\left\{\vartheta_1,\ldots,\vartheta_G\right\}$, and $\text{\textsubtilde{$\nu$}}=\left\{\nu_1,\ldots,\nu_G\right\}$), respectively .

\begin{prop}
Given $\text{\textsubtilde{$\pi$}}$, $\text{\textsubtilde{$\boldsymbol{\vartheta}$}}$, and $\text{\textsubtilde{$\nu$}}$,  
if $h_t\left(y|\bx;\bxi_1,\zeta_1\right)=\cdots=h_t\left(y|\bx;\bxi_G,\zeta_G\right)=h_t\left(y|\bx;\bxi,\zeta\right)$, then models \eqref{eq:linear $t$ CWM} and \eqref{eq:$t$ FMD} generate the same posterior probabilities.
\label{lem:CWM vs. FMD}
\end{prop}
\begin{pf}
If the component conditional densities do not depend on $\Omega_g$, then the posterior probabilities for the linear $t$ CWM in \eqref{eq:linear $t$ CWM} can be written as 
\begin{displaymath}
\tau_{ng}=\frac{\pi_g  \cancel{h_t\left(y_n|\bx_n;\bxi,\zeta\right)}h_{t_d}\left(\bx_n;\bvartheta_g,\nu_g\right)}{
\displaystyle\sum_{j=1}^G\pi_j\cancel{ h_t\left(y_n|\bx_n;\bxi,\zeta\right)}h_{t_d}\left(\bx_n;\bvartheta_j,\nu_j\right)}=\frac{\pi_g h_{t_d}\left(\bx_n;\bvartheta_g,\nu_g\right)}{
\displaystyle\sum_{j=1}^G\pi_jh_{t_d}\left(\bx_n;\bvartheta_j,\nu_j\right)},
\end{displaymath}
which correspond to the posterior probabilities for model \eqref{eq:$t$ FMD}.\qed
\end{pf}
\begin{prop}
Given $\text{\textsubtilde{$\pi$}}$, $\text{\textsubtilde{$\boldsymbol{\xi}$}}$, and $\text{\textsubtilde{$\zeta$}}$,  
if $h_{t_d}\left(\bx;\bvartheta_1,\nu_1\right)=\cdots=h_{t_d}\left(\bx;\bvartheta_G,\nu_G\right)=h_{t_d}\left(\bx;\bvartheta,\nu\right)$, then models \eqref{eq:linear $t$ CWM} and \eqref{eq:linear $t$ FMR} generate the same posterior probabilities.
\label{lem:CWM vs. FMR}
\end{prop}
\begin{pf}
If the component marginal densities do not depend on $\Omega_g$, then the posterior probabilities for the linear $t$ CWM in \eqref{eq:linear $t$ CWM} can be written as 
\begin{displaymath}
\tau_{ng}=\frac{\pi_g  h_t\left(y_n|\bx_n;\bxi_g,\zeta_g\right)\cancel{h_{t_d}\left(\bx_n;\bvartheta,\nu\right)}}{
\displaystyle\sum_{j=1}^G\pi_jh_t\left(y_n|\bx_n;\bxi_j,\zeta_j\right)\cancel{h_{t_d}\left(\bx_n;\bvartheta,\nu\right)}}=\frac{\pi_g h_t\left(y_n|\bx_n;\bxi_g,\zeta_g\right)}{
\displaystyle\sum_{j=1}^G\pi_jh_t\left(y_n|\bx_n;\bxi_j,\zeta_j\right)},
\end{displaymath}
which correspond to the posterior probabilities for model \eqref{eq:linear $t$ FMR}.\qed
\end{pf}
Note that the results in Proposition~\ref{lem:CWM vs. FMD} and \ref{lem:CWM vs. FMR} are not restricted to the $t$ distribution; in fact, they can be easily extended to the general CWM in \eqref{eq:general cwm}.
Further, some results about the relation between linear Gaussian (or $t$) CWMs and finite mixture of regressions are given in \cite{Ingr:Mino:Vitt:Loca:2012}.
Finally, it is important to underline that up to now there are no theoretical results on the identifiability for linear CWMs; however, since they can be seen as mixture models with random covariates, the results in \citet[][Section~3, Model 2.a]{Henn:Iden:2000} can apply.

\section{The family of linear CWMs}
\label{sec:The parsimonious Student-t LCWMs}

This section introduces the novel family of mixture models obtained from the linear $t$ CWM.
In \eqref{eq:linear $t$ CWM}, let us consider: 
\begin{itemize}
	\item component conditional densities $h_t$ having the same parameters for all $\Omega_g$,
	\item component marginal densities $h_{t_d}$ having the same parameters for all $\Omega_g$,
	\item degrees of freedom $\zeta_g$ tending to infinity for each $\Omega_g$, and 
	\item degrees of freedom $\nu_g$ tending to infinity for each $\Omega_g$.
\end{itemize}
By combining such constraints, we obtain twelve parsimonious and easily interpretable linear CWMs that are appropriate for describing various practical situations; they are schematically presented in \tablename~\ref{tab:parsimonious models} along with the number of parameters characterizing each component of the CW decomposition.
For instance, if $\nu_g,\zeta_g\rightarrow\infty$ for each $\Omega_g$, we are assuming a normal distribution for the component conditional and marginal densities; furthermore, we can assume different linear models (in terms of $\bbeta_g$ and $\sigma^2_g$) in each cluster while keeping the density of $\boldsymbol{X}$ equal between clusters.
From a notational viewpoint, this leads to a linear CWM that we have simply denoted as $NN$-EV: the first two letters represent the distribution of $\bX|\Omega_g$ and $Y|\bX,\Omega_g$ ($N\equiv$Normal and $t\equiv$$t$), respectively, while the second two denote the distribution constraint between clusters (E$\equiv$Equal and V$\equiv$Variable) for $\bX|\Omega_g$ and $Y|\bX,\Omega_g$, respectively.
\setlength\arrayrulewidth{1.4pt}
\setlength{\extrarowheight}{3pt}
\begin{sidewaystable}
\centering
\begin{tabular}{cccccccccc}
\toprule
Model & \multicolumn{2}{c}{$\bX|\Omega_g$} & \multicolumn{2}{c}{$Y|\bx,\Omega_g$} & \multicolumn{5}{c}{Number of free parameters} \\
Identifier & Density & Constraint & Density & Constraint & $\bX$ & & $Y|\bx$ & & weights\\
\midrule
$tt$-VV        & $t$  & Variable  & $t$  & Variable  &  $G\left(d+\frac{d\left(d+1\right)}{2}+1\right)$ & + &  $G\left(d+3\right)$ & + & $G-1$\\
$tt$-VE        & $t$  & Variable  & $t$  & Equal      &  $G\left(d+\frac{d\left(d+1\right)}{2}+1\right)$ & + &  $d+3$ & + & $G-1$\\
$tt$-EV        & $t$  & Equal      & $t$  & Variable  &  $d+\frac{d\left(d+1\right)}{2}+1$ & + &  $G\left(d+3\right)$ & + & $G-1$\\
$NN$-VV        & Normal     & Variable  & Normal     & Variable  &  $G\left(d+\frac{d\left(d+1\right)}{2}\right)$ & + &  $G\left(d+2\right)$ & + & $G-1$\\
$NN$-VE        & Normal     & Variable  & Normal     & Equal      &  $G\left(d+\frac{d\left(d+1\right)}{2}\right)$ & + &  $d+2$ & + & $G-1$\\
$NN$-EV        & Normal     & Equal      & Normal     & Variable  &  $d+\frac{d\left(d+1\right)}{2}$ & + &  $G\left(d+2\right)$ & + & $G-1$\\
$tN$-VV        & $t$  & Variable  & Normal     & Variable  &  $G\left(d+\frac{d\left(d+1\right)}{2}+1\right)$ & + &  $G\left(d+2\right)$ & + & $G-1$\\
$tN$-VE        & $t$  & Variable  & Normal     & Equal      &  $G\left(d+\frac{d\left(d+1\right)}{2}+1\right)$ & + &  $d+2$ & + & $G-1$\\
$tN$-EV        & $t$  & Equal      & Normal     & Variable  &  $d+\frac{d\left(d+1\right)}{2}+1$ & + &  $G\left(d+2\right)$ & + & $G-1$\\
$Nt$-VV        & Normal     & Variable  & $t$  & Variable  &  $G\left(d+\frac{d\left(d+1\right)}{2}\right)$ & + &  $G\left(d+3\right)$ & + & $G-1$\\
$Nt$-VE        & Normal     & Variable  & $t$  & Equal      &  $G\left(d+\frac{d\left(d+1\right)}{2}\right)$ & + &  $d+3$ & + & $G-1$\\
$Nt$-EV        & Normal     & Equal      & $t$  & Variable  &  $d+\frac{d\left(d+1\right)}{2}$ & + &  $G\left(d+3\right)$ & + & $G-1$\\
\bottomrule
\end{tabular}
\caption{Overview of linear CWMs. 
In ``model identifier'', the first and second letters represent, respectively, the density of $\bX|\Omega_g$ and $Y|\bx,\Omega_g$ (here $N\equiv$Normal), while the third and fourth letters indicate, respectively, if $h_{t_d}\left(\bx;\bvartheta_g,\nu_g\right)$ and $h_t\left(y|\bx;\bxi_g,\zeta_g\right)$ are assumed to be Equal$\equiv$E or Variable$\equiv$V between groups. 
} 
\label{tab:parsimonious models}
\end{sidewaystable}

Only two of the models given in \tablename~\ref{tab:parsimonious models}, $NN$-VV and $tt$-VV, have been developed previously; the former corresponds to the linear Gaussian CWM of \citet{Gers:Nonl:1997}, while the latter coincides with the linear $t$ CWM in \citet{Ingr:Mino:Vitt:Loca:2012}.
Furthermore, in principle there are sixteen models arising from the combination of the aforementioned constraints; nevertheless, four of them -- those which should be denoted as EE -- do not make sense. 
Indeed, they lead to a single cluster regardless of the value of $G$.
Finally, we remark that when $G=1$, it results $\text{VV}\equiv\text{VE}\equiv\text{EV}$ regardless of the chosen distribution. 

\section{Estimation via the EM algorithm}
\label{sec:Estimation via the EM algorithm}

The EM algorithm \citep{Demp:Lair:Rubi:Maxi:1977} is the standard tool for maximum likelihood (ML) estimation of the parameters for mixture models.
This section describes the EM algorithm for the most general model $tt$-VV.
Details for all the other models are given in \ref{app:Constraints for parsimonious models}.

In the EM framework, the generic observation $\left(y_n,\boldsymbol{x}_n'\right)'$ is viewed as being incomplete; its complete counterpart is given by $\left(y_n,\boldsymbol{x}_n',\boldsymbol{z}_n',u_n,v_n\right)'$, where $\boldsymbol{z}_n$ is the component-label vector in which $z_{ng}=1$ if $\left(y_n,\boldsymbol{x}_n'\right)'$ comes from the $g$th component ($z_{ng}=0$ otherwise), while $u_n$ and $v_n$ arise from the standard theory of the (multivariate) $t$ distribution according to which
\begin{eqnarray}
Y_n\left|\bx_n,v_n,z_{ng}=1\right.&\stackrel{\text{ind.}}{\sim} &\text{N}\left(\mu\left(\bx_n;\bbeta_g\right),\frac{\sigma^2_g}{v_n}\right)\label{eq:distribution Y|x}\\
V_n\left|z_{ng}=1\right.&\stackrel{\text{i.i.d.}}{\sim}& \text{Gamma}\left(\frac{\zeta_g}{2},\frac{\zeta_g}{2}\right)\label{eq:distribution V},
\end{eqnarray}
for $n=1,\ldots,N$, and
\begin{eqnarray}
\bX_n\left|u_n,z_{ng}=1\right.&\stackrel{\text{ind.}}{\sim} &\text{N}\left(\bmu_g,\frac{\bSigma_g}{u_n}\right)\label{eq:distribution X}\\
U_n\left|z_{ng}=1\right.&\stackrel{\text{i.i.d.}}{\sim}& \text{Gamma}\left(\frac{\nu_g}{2},\frac{\nu_g}{2}\right)\label{eq:distribution U},
\end{eqnarray}
for $n=1,\ldots,N$.
Because of the conditional structure of the complete-data model given by distributions \eqref{eq:distribution Y|x}, \eqref{eq:distribution V}, \eqref{eq:distribution X}, and \eqref{eq:distribution U}, the complete-data log-likelihood can be decomposed as
\begin{equation}
l_c\left(\text{\textsubtilde{$\boldsymbol{\psi}$}}\right)=l_{1c}\left(\text{\textsubtilde{$\pi$}}\right)+l_{2c}\left(\text{\textsubtilde{$\boldsymbol{\xi}$}}\right)+l_{3c}\left(\text{\textsubtilde{$\zeta$}}\right)+l_{4c}\left(\text{\textsubtilde{$\boldsymbol{\vartheta}$}}\right)+l_{5c}\left(\text{\textsubtilde{$\nu$}}\right),
\label{eq:complete-data log-likelihood}
\end{equation}
where
\begin{displaymath}
l_{1c}\left(\text{\textsubtilde{$\pi$}}\right)=\sum_{n=1}^N\sum_{g=1}^G z_{ng}\ln \pi_g,
\end{displaymath}
\begin{displaymath}
l_{2c}\left(\text{\textsubtilde{$\boldsymbol{\xi}$}}\right)=\frac{1}{2}\sum_{n=1}^N\sum_{g=1}^G z_{ng}\Bigl\{-\ln\left(2\pi\right)+\ln v_n-\ln\sigma^2_g - v_n\delta\left[y_n,\mu\left(\bx_n;\bbeta_g\right);\sigma^2_g\right]\Bigr\},
\end{displaymath}
\begin{displaymath}
l_{3c}\left(\text{\textsubtilde{$\zeta$}}\right) = \sum_{n=1}^N\sum_{g=1}^G z_{ng}\left[-\ln\Gamma\left(\frac{\zeta_g}{2}\right) + \frac{\zeta_g}{2}\ln\frac{\zeta_g}{2} + \frac{\zeta_g}{2} \left(\ln v_n - v_n\right) - \ln v_n\right],
\end{displaymath}
\begin{displaymath}
l_{4c}\left(\text{\textsubtilde{$\boldsymbol{\vartheta}$}}\right) = \frac{1}{2}\sum_{n=1}^N\sum_{g=1}^G z_{ng}\left[-d\ln\left(2\pi\right)+d\ln u_n-\ln\left|\bSigma_g\right| - u_n\delta\left(\bx_n,\bmu_g;\bSigma_g\right)\right]
\end{displaymath}
and
\begin{displaymath}
l_{5c}\left(\text{\textsubtilde{$\nu$}}\right) = \sum_{n=1}^N\sum_{g=1}^G  z_{ng} \left[-\ln\Gamma\left(\frac{\nu_g}{2}\right) + \frac{\nu_g}{2} \ln\frac{\nu_g}{2} + \frac{\nu_g}{2} \left(\ln u_n - u_n\right) - \ln u_n\right]. 
\end{displaymath}

\subsection[E-step]{E-step}
\label{subsec:E-step}

The E-step, on the $\left(k+1\right)$th iteration, requires the calculation of
\begin{equation}
Q\left(\text{\textsubtilde{$\boldsymbol{\psi}$}};\text{\textsubtilde{$\boldsymbol{\psi}$}}^{\left(k\right)}\right)=E_{\text{\textsubtilde{$\boldsymbol{\psi}$}}^{\left(k\right)}}\left[l_c\left(\text{\textsubtilde{$\boldsymbol{\psi}$}}\right)\left|\left(y_1,\bx_1'\right)',\ldots,\left(y_n,\bx_n'\right)'\right.\right].
\label{eq:Q}
\end{equation}
In order to do this, we need to calculate $E_{\text{\textsubtilde{$\boldsymbol{\psi}$}}^{\left(k\right)}}\left(Z_{ng}\left|y_n,\bx_n\right.\right)$, $E_{\text{\textsubtilde{$\boldsymbol{\psi}$}}^{\left(k\right)}}\left(V_n\left|y_n,\bx_n,\boldsymbol{z}_n\right.\right)$, $E_{\text{\textsubtilde{$\boldsymbol{\psi}$}}^{\left(k\right)}}\left(\widetilde{V}_n\left|y_n,\bx_n,\boldsymbol{z}_n\right.\right)$, $E_{\text{\textsubtilde{$\boldsymbol{\psi}$}}^{\left(k\right)}}\left(U_n\left|\bx_n,\boldsymbol{z}_n\right.\right)$, and $E_{\text{\textsubtilde{$\boldsymbol{\psi}$}}^{\left(k\right)}}\left(\widetilde{U}_n\left|\bx_n,\boldsymbol{z}_n\right.\right)$, for $n=1,\ldots,N$ and $g=1,\ldots,G$, where $\widetilde{U}_n=\ln U_n$ and $\widetilde{V}_n=\ln V_n$.
It follows that 
\begin{eqnarray}
E_{\text{\textsubtilde{$\boldsymbol{\psi}$}}^{\left(k\right)}}\left(Z_{ng}\left|y_n,\bx_n\right.\right)&=&\tau_{ng}^{\left(k\right)}\nonumber\\
&=&\frac{\pi_g^{\left(k\right)} h_t\left(y_n|\bx_n;\boldsymbol{\xi}_g^{\left(k\right)},\zeta_g^{\left(k\right)}\right)h_{t_d}\left(\bx_n;\boldsymbol{\vartheta}_g^{\left(k\right)},\nu_g^{\left(k\right)}\right)}{p\left(y_n,\bx_n;\text{\textsubtilde{$\boldsymbol{\psi}$}}^{\left(k\right)}\right)},
\label{eq:EZ}
\end{eqnarray}
\begin{eqnarray}
E_{\text{\textsubtilde{$\boldsymbol{\psi}$}}^{\left(k\right)}}\left(V_n\left|y_n,\bx_n,z_{ng}=1\right.\right)&=&v_{ng}^{\left(k\right)}\nonumber\\
&=&\frac{ \zeta_{g}^{\left(k\right)}+1 }{\zeta_g^{\left(k\right)} + \delta\left[y_n,\mu\left(\bx_n;\bbeta_g^{\left(k\right)}\right);\sigma^{2\left(r\right)}_g\right]
}
\label{eq:EV}
\end{eqnarray}
and
\begin{eqnarray}
E_{\text{\textsubtilde{$\boldsymbol{\psi}$}}^{\left(k\right)}}\left(U_n\left|y_n,\bx_n,z_{ng}=1\right.\right)&=&u_{ng}^{\left(k\right)}\nonumber\\
&=&\frac{\nu_{g}^{\left(k\right)}+d}{\nu_g^{\left(k\right)} + \delta\left(\bx_n,\bmu_g^{\left(k\right)};\bSigma_g^{\left(k\right)}\right)},
\label{eq:EU}
\end{eqnarray}
where the expectations are affected (see the subscript) using the current fit $\text{\textsubtilde{$\boldsymbol{\psi}$}}^{\left(k\right)}$ for $\text{\textsubtilde{$\boldsymbol{\psi}$}}$ ($n=1,\ldots,N$ and $g=1,\ldots,G$).
Regarding the last two expectations, from the standard theory on the gamma distribution, we have that
\begin{eqnarray}
E_{\text{\textsubtilde{$\boldsymbol{\psi}$}}^{\left(k\right)}}\left(\widetilde{V}_n\left|y_n,\bx_n,z_{ng}=1\right.\right)
&=&\widetilde{v}_{ng}^{\left(k\right)} \nonumber\\
&=& 
\ln v_{ng}^{\left(k\right)} + \psi\left(\frac{\zeta_{g}^{\left(k\right)} + 1}{2}\right) - \ln \left(\frac{\zeta_{g}^{\left(k\right)} + 1}{2}\right) 
\label{eq:ElnV}
\end{eqnarray}
and
\begin{eqnarray}
E_{\text{\textsubtilde{$\boldsymbol{\psi}$}}^{\left(k\right)}}\left(\widetilde{U}_n\left|\bx_n,z_{ng}=1\right.\right)
&=&\widetilde{u}_{ng}^{\left(k\right)} \nonumber\\
&=& 
\ln u_{ng}^{\left(k\right)} + \psi\left(\frac{\nu_{g}^{\left(k\right)} + d}{2}\right) - \ln \left(\frac{\nu_{g}^{\left(k\right)} + d}{2}\right),  
\label{eq:ElnU}
\end{eqnarray}
where $\psi\left(s\right)=\left[\partial\Gamma\left(s\right)/\partial s\right]/\Gamma\left(s\right)$ is the Digamma function.

Using the results from \eqref{eq:EZ} to \eqref{eq:ElnV} to calculate \eqref{eq:Q}, we have that 
\begin{equation}
Q\left(\text{\textsubtilde{$\boldsymbol{\psi}$}};\text{\textsubtilde{$\boldsymbol{\psi}$}}^{\left(k\right)}\right)=Q_1\left(\text{\textsubtilde{$\pi$}};\text{\textsubtilde{$\boldsymbol{\psi}$}}^{\left(k\right)}\right)+Q_2\left(\text{\textsubtilde{$\boldsymbol{\xi}$}};\text{\textsubtilde{$\boldsymbol{\psi}$}}^{\left(k\right)}\right)+Q_3\left(\text{\textsubtilde{$\zeta$}};\text{\textsubtilde{$\boldsymbol{\psi}$}}^{\left(k\right)}\right)+Q_4\left(\text{\textsubtilde{$\boldsymbol{\vartheta}$}};\text{\textsubtilde{$\boldsymbol{\psi}$}}^{\left(k\right)}\right)+Q_5\left(\text{\textsubtilde{$\nu$}};\text{\textsubtilde{$\boldsymbol{\psi}$}}^{\left(k\right)}\right),
\label{eq:decompositionQ1}
\end{equation}
where
\begin{equation}
Q_1\left(\text{\textsubtilde{$\pi$}};\text{\textsubtilde{$\boldsymbol{\psi}$}}^{\left(k\right)}\right)=\sum_{n=1}^N\sum_{g=1}^G\tau_{ng}^{\left(k\right)}\ln \pi_g,
\label{eq:Q1}
\end{equation}
\begin{equation}
Q_2\left(\text{\textsubtilde{$\boldsymbol{\xi}$}};\text{\textsubtilde{$\boldsymbol{\psi}$}}^{\left(k\right)}\right)=\sum_{n=1}^N\sum_{g=1}^G\tau_{ng}^{\left(k\right)}Q_{2n}\left(\boldsymbol{\xi}_g;\text{\textsubtilde{$\boldsymbol{\psi}$}}^{\left(k\right)}\right),
\label{eq:Q2}
\end{equation}
\begin{equation}
Q_3\left(\text{\textsubtilde{$\zeta$}};\text{\textsubtilde{$\boldsymbol{\psi}$}}^{\left(k\right)}\right)=\sum_{n=1}^N\sum_{g=1}^G\tau_{ng}^{\left(k\right)}Q_{3n}\left(\zeta_g;\text{\textsubtilde{$\boldsymbol{\psi}$}}^{\left(k\right)}\right),
\label{eq:Q3}
\end{equation}
\begin{equation}
Q_4\left(\text{\textsubtilde{$\boldsymbol{\vartheta}$}};\text{\textsubtilde{$\boldsymbol{\psi}$}}^{\left(k\right)}\right)=\sum_{n=1}^N\sum_{g=1}^G\tau_{ng}^{\left(k\right)}Q_{4n}\left(\boldsymbol{\vartheta}_g;\text{\textsubtilde{$\boldsymbol{\psi}$}}^{\left(k\right)}\right)
\label{eq:Q4}
\end{equation}
and
\begin{equation}
Q_5\left(\text{\textsubtilde{$\nu$}};\text{\textsubtilde{$\boldsymbol{\psi}$}}^{\left(k\right)}\right)=\sum_{n=1}^N\sum_{g=1}^G\tau_{ng}^{\left(k\right)}Q_{5n}\left(\nu_g;\text{\textsubtilde{$\boldsymbol{\psi}$}}^{\left(k\right)}\right),
\label{eq:Q5}
\end{equation}
with 
\begin{displaymath}
Q_{2n}\left(\boldsymbol{\xi}_g;\text{\textsubtilde{$\boldsymbol{\psi}$}}^{\left(k\right)}\right)=\frac{1}{2}\left\{-\ln\left(2\pi\right)+\widetilde{v}_{ng}^{\left(k\right)}-\ln\sigma^2_g-v_{ng}\delta\left[y_n,\mu\left(\bx_n;\bbeta_g\right);\sigma^{2}_g\right]\right\}
\end{displaymath}
and
\begin{displaymath}
Q_{4n}\left(\boldsymbol{\vartheta}_g;\text{\textsubtilde{$\boldsymbol{\psi}$}}^{\left(k\right)}\right)=\frac{1}{2}\left[-d\ln\left(2\pi\right)+d\widetilde{u}_{ng}^{\left(k\right)}-\ln\left|\bSigma_g\right|-u_{ng}\delta\left(\bx_n,\bmu_g;\bSigma_g\right)\right],
\end{displaymath}
and where, on ignoring terms not involving $\zeta_g$ and $\nu_g$, respectively,
\begin{displaymath}
Q_{3n}\left(\zeta_g;\text{\textsubtilde{$\boldsymbol{\psi}$}}^{\left(k\right)}\right)=-\ln\Gamma\left(\frac{\zeta_g}{2}\right)+\frac{\zeta_g}{2}\ln \frac{\zeta_g}{2}+\frac{\zeta_g}{2}\left[\widetilde{v}_{ng}^{\left(k\right)}-\ln v_{ng}^{\left(k\right)}+\sum_{n=1}^N\left(\ln v_{ng}^{\left(k\right)}-v_{ng}^{\left(k\right)} \right)\right]
\end{displaymath}
and
\begin{displaymath}
Q_{5n}\left(\nu_g;\text{\textsubtilde{$\boldsymbol{\psi}$}}^{\left(k\right)}\right)=-\ln\Gamma\left(\frac{\nu_g}{2}\right)+\frac{\nu_g}{2}\ln \frac{\nu_g}{2}+\frac{\nu_g}{2}\left[\widetilde{u}_{ng}^{\left(k\right)}-\ln u_{ng}^{\left(k\right)}+\sum_{n=1}^N\left(\ln u_{ng}^{\left(k\right)}-u_{ng}^{\left(k\right)} \right)\right].
\end{displaymath}

\subsection[M-step]{M-step}
\label{subsec:M-step}

On the M-step, at the $\left(k+1\right)$th iteration, it follows from \eqref{eq:decompositionQ1} that $\text{\textsubtilde{$\pi$}}^{\left(k+1\right)}$, $\text{\textsubtilde{$\boldsymbol{\xi}$}}^{\left(k+1\right)}$, $\text{\textsubtilde{$\zeta$}}^{\left(k+1\right)}$, $\text{\textsubtilde{$\boldsymbol{\vartheta}$}}^{\left(k+1\right)}$, and $\text{\textsubtilde{$\nu$}}^{\left(k+1\right)}$ can be computed independently of each other, by separate consideration of \eqref{eq:Q1}, \eqref{eq:Q2}, \eqref{eq:Q3}, \eqref{eq:Q4}, and \eqref{eq:Q5}, respectively.
The solutions for $\pi_g^{\left(k+1\right)}$, $\boldsymbol{\xi}_g^{\left(k+1\right)}$, and 
$\boldsymbol{\vartheta}_g^{\left(k+1\right)}$ exist in closed form.
Only the updates $\zeta_g^{\left(k+1\right)}$ and $\nu_g^{\left(k+1\right)}$ need to be computed iteratively.

The updated estimates of the mixture weights are
\begin{equation}
\pi_g^{\left(k+1\right)}=\sum_{n=1}^N \tau_{ng}^{\left(k\right)}\Big/ n,
\label{eq:updated weights}
\end{equation}
while those of $\boldsymbol{\vartheta}_g$, $g=1,\ldots,G$, result 
\begin{equation}
\bmu_g^{\left(k+1\right)} = \displaystyle\sum_{n=1}^N\tau_{ng}^{\left(k\right)}u_{ng}^{\left(k\right)}\bx_n\Big/\displaystyle\sum_{n=1}^N\tau_{ng}^{\left(k\right)}u_{ng}^{\left(k\right)}\label{eq:updated mean vector for X} 
\end{equation}
and
\begin{equation}
\bSigma_g^{\left(k+1\right)} = \displaystyle\sum_{n=1}^N\tau_{ng}^{\left(k\right)}u_{ng}^{\left(k\right)}\left(\bx_n-\bmu_g^{\left(k+1\right)}\right)\left(\bx_n-\bmu_g^{\left(k+1\right)}\right)'\Big/\displaystyle\sum_{n=1}^N\tau_{ng}^{\left(k\right)}u_{ng}^{\left(k\right)},
\label{eq:updated covariance matrix for X}
\end{equation}
where, as motivated for example in 
\citet{Shoa:Patt:2002}, 
the true denominator $\sum_n\tau_{ng}^{\left(k\right)}$ of \eqref{eq:updated covariance matrix for X} has been changed to yield a significantly faster convergence for the EM algorithm.


Regarding the updated estimates of $\boldsymbol{\xi}_g$, $g=1,\ldots,G$, maximization of \eqref{eq:Q2}, after some algebra, yields
\begin{eqnarray}
\bbeta_{1g}^{\left(k+1\right)}&=&\left(\frac{\displaystyle\sum_{n=1}^N\tau_{ng}^{\left(k\right)}v_{ng}^{\left(k\right)}\bx_n\bx_n'}{\displaystyle\sum_{n=1}^N\tau_{ng}^{\left(k\right)}v_{ng}^{\left(k\right)}}-\frac{\displaystyle\sum_{n=1}^N\tau_{ng}^{\left(k\right)}v_{ng}^{\left(k\right)}\bx_n}{\displaystyle\sum_{n=1}^N\tau_{ng}^{\left(k\right)}v_{ng}^{\left(k\right)}
}\frac{\displaystyle\sum_{n=1}^N\tau_{ng}^{\left(k\right)}v_{ng}^{\left(k\right)}\bx_n'}{\displaystyle\sum_{n=1}^N\tau_{ng}^{\left(k\right)}v_{ng}^{\left(k\right)}
}\right)^{-1}
\cdot\nonumber\\
&&\cdot \left(
\frac{\displaystyle\sum_{n=1}^N\tau_{ng}^{\left(k\right)}v_{ng}^{\left(k\right)}y_n\bx_n}{\displaystyle\sum_{n=1}^N\tau_{ng}^{\left(k\right)}v_{ng}^{\left(k\right)}}
-
\frac{\displaystyle\sum_{n=1}^N\tau_{ng}^{\left(k\right)}v_{ng}^{\left(k\right)}y_n}{\displaystyle\sum_{n=1}^N\tau_{ng}^{\left(k\right)}v_{ng}^{\left(k\right)}
}\frac{\displaystyle\sum_{n=1}^N\tau_{ng}^{\left(k\right)}v_{ng}^{\left(k\right)}\bx_n}{\displaystyle\sum_{n=1}^N\tau_{ng}^{\left(k\right)}v_{ng}^{\left(k\right)}
}\right),
\label{eq:updated beta1}
\end{eqnarray}
\begin{equation}
\beta_{0g}^{\left(k+1\right)} = \displaystyle \frac{\displaystyle\sum_{n=1}^N\tau_{ng}^{\left(k\right)}v_{ng}^{\left(k\right)}y_n}{
\displaystyle\sum_{n=1}^N\tau_{ng}^{\left(k\right)}v_{ng}^{\left(k\right)}}
-\bbeta_{1g}^{\left(k+1\right)'}
\frac{\displaystyle\sum_{n=1}^N\tau_{ng}^{\left(k\right)}v_{ng}^{\left(k\right)}\bx_n}{\displaystyle\sum_{n=1}^N\tau_{ng}^{\left(k\right)}v_{ng}^{\left(k\right)}
}\label{eq:updated beta0}
\end{equation}
and
\begin{equation}
\sigma^{2\left(k+1\right)}_g=
\displaystyle\displaystyle\sum_{n=1}^N\tau_{ng}^{\left(k\right)}v_{ng}^{\left(k\right)}\left[y_n-\left(\beta_{0g}^{\left(k+1\right)}+\boldsymbol{\beta}_{1g}^{\left(k+1\right)'}\bx_n\right)\right]^2\Big/\displaystyle\sum_{n=1}^N\tau_{ng}^{\left(k\right)}v_{ng}^{\left(k\right)},\label{eq:updated conditioned variances}
\end{equation}
where the denominator of \eqref{eq:updated conditioned variances} has been modified in line with what was explained for equation \eqref{eq:updated covariance matrix for X}.  

As said before, because we are acting in the most general case in which the degrees of freedom $\zeta_g$ and $\nu_g$ are inferred from the data, we need to numerically solve the equations
\begin{equation}
\sum_{n=1}^N\frac{\partial}{\partial \zeta_g}Q_{3n}\left(\zeta_g;\text{\textsubtilde{$\boldsymbol{\psi}$}}^{\left(k\right)}\right)=0
\label{eq:numerical eq for zeta}
\end{equation}
and
\begin{equation}
\sum_{n=1}^N\frac{\partial}{\partial \nu_g}Q_{5n}\left(\nu_g;\text{\textsubtilde{$\boldsymbol{\psi}$}}^{\left(k\right)}\right)=0,
\label{eq:numerical eq for nu}
\end{equation}
which correspond to finding $\zeta_g^{\left(k+1\right)}$ and $\nu_g^{\left(k+1\right)}$ as the respective solutions of
\begin{eqnarray}
&-\psi\left(\displaystyle\frac{\zeta_g}{2}\right) + \ln\displaystyle\frac{\zeta_g}{2}+ 1  +\frac{1}{N_g^{\left(k\right)}} \displaystyle\sum_{n=1}^N \tau_{ng}^{\left(k\right)}\left(\ln v_{ng}^{\left(k\right)} - v_{ng}^{\left(k\right)}\right)+&\nonumber\\
&  \psi\left(\displaystyle\frac{\zeta_g^{\left(k\right)}+1}{2}\right)-\ln\left(\displaystyle\frac{\zeta_g^{\left(k\right)}+1}{2}\right)  = 0& \label{eq:zeta}
\end{eqnarray}
and 
\begin{eqnarray}
&-\psi\left(\displaystyle\frac{\nu_g}{2}\right) + \ln\displaystyle\frac{\nu_g}{2}+ 1 + \displaystyle\frac{1}{N_g^{\left(k\right)}} \displaystyle\sum_{n=1}^N \tau_{ng}^{\left(k\right)} \left(\ln u_{ng}^{\left(k\right)}- u_{ng}^{\left(k\right)} \right) + & \nonumber\\
&  \psi\left(\displaystyle\frac{\nu_g^{\left(k\right)}+d}{2}\right)-\ln\left(\displaystyle\frac{\nu_g^{\left(k\right)}+d}{2}\right)=0, & \label{eq:nu}
\end{eqnarray}
where $N_g^{\left(k\right)}=\sum_n\tau_{ng}^{\left(k\right)}$, $g=1,\ldots,G$.

\section{Computational issues and partition evaluation}
\label{sec:Computational issues and partitions evaluation}

This section presents some issues concerning practical implementation of the EM algorithm described in Section~\ref{sec:Estimation via the EM algorithm} (see also \ref{app:Constraints for parsimonious models}). 

\subsection{Estimating the degrees of freedom}
\label{subsec:Estimating the degrees of freedom}

Code for all of the analyses presented herein was written in the R computing environment \citep{R} and a numerical search for the estimates of the degrees of freedom was carried out using the \texttt{uniroot} command in the \texttt{stats} package. 
This command is based on the Fortran subroutine \texttt{zeroin} described by \citet{Bren:Algo:1973}. 
In order to expedite convergence, the range of values for $\nu_g$, $\zeta_g$, $\nu$, and $\zeta$ was restricted to $\left(2,200\right]$. 
Previous work in the context of model-based clustering \citep[see][]{Andr:McNi:Exte:2011} and some experiments whose results are not reported here suggest that these restrictions do not hamper classification performance and show that the upper limit of 200 does not thwart the recovery of an underlying normal structure.

\subsection{EM initialization}
\label{subsec:EM initialization}

It is well known that the choice of starting values represents an important issue in the EM algorithm.
The standard initialization consists of selecting a value for $\text{\textsubtilde{$\bpsi$}}^{(0)}$ \citep[see, e.g.,][]{Bagn:Punz:Fine:2013}.
An alternative approach, more natural in the authors' opinion, is to specify a value for $\boldsymbol{z}_n^{(0)}$, $n=1,\ldots,N$ \citep[see][p.~54]{McLa:Peel:fini:2000}.   
Within this approach, and due to the structure of our family of linear CWMs, we propose a random-hierarchical initialization procedure that helps in obtaining the natural ranking among the likelihoods.

\begin{figure}[!ht]
\centering
\resizebox{0.7\textwidth}{!}{
\includegraphics{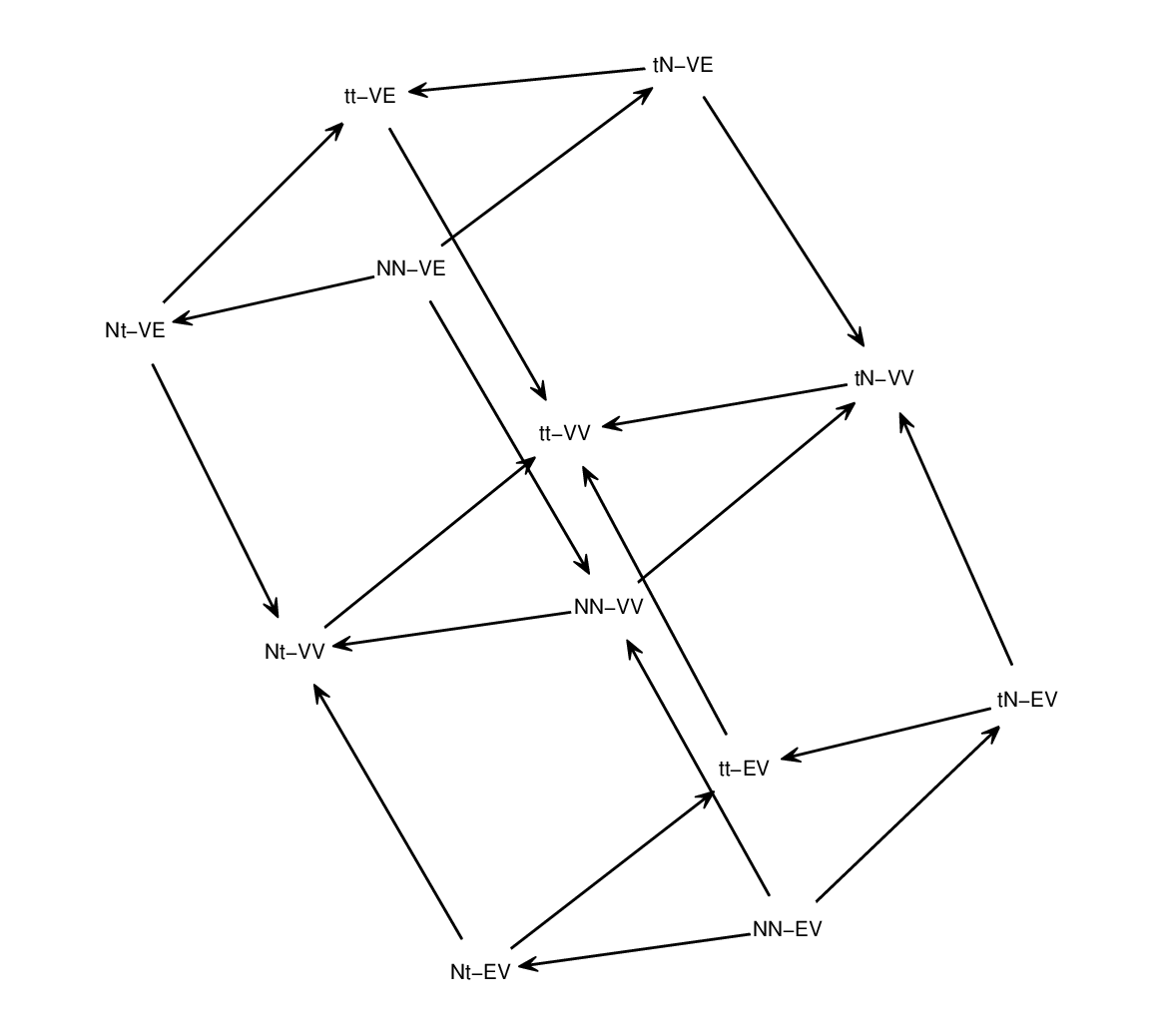} 
}
\caption{
Relationships among the models in the hierarchical initialization strategy.
Arrows are oriented from the model used for initialization to the model to be estimated.   
}
\label{fig:network}
\end{figure}

For a fixed $G$, we start by considering $NN$-VE and $NN$-EV, because the former is nested in all of the VE-models, the latter is nested in all of the EV models, and both are nested in all of the VV-models.
For $NN$-VE and $NN$-EV only, a random initialization is repeated 10 times, from different random positions, and the solution maximizing the likelihood among these 10 runs is selected.
Note that, as underlined by \citet{Andr:McNi:Sube:Mode:2011}, mixtures based on the multivariate $t$ distribution are more sensitive to bad starting values than their Gaussian counterparts.
Thus, by considering random initialization only for the above models of type $NN$, we prevent the possible failure of the algorithm due to poor starting values for models of type $Nt$, $tN$, and $tt$. 
In each run, the $N$ vectors $\boldsymbol{z}_n^{(0)}$ are randomly drawn from a multinomial distribution with probabilities $\left(1/G,\ldots,1/G\right)$.
Once the EM-estimates $\widehat{\tau}_{ng}^{\text{$NN$-VE}}$ and $\widehat{\tau}_{ng}^{\text{$NN$-EV}}$ of the posterior probabilities have been obtained for these models, we can compute the maximum \textit{a posteriori} (MAP) classification, say $\text{MAP}\left(\widehat{\tau}_{ng}^{\text{$NN$-VE}}\right)=\widehat{z}_{ng}^{\text{$NN$-VE}}$ and $\text{MAP}\left(\widehat{\tau}_{ng}^{\text{$NN$-EV}}\right)=\widehat{z}_{ng}^{\text{$NN$-EV}}$, where 
\[
\text{MAP}\left(\widehat{\tau}_{ng}\right) = \widehat{z}_{ng} = \left\{
\begin{array}{ll}
 1 & \text{if $\displaystyle\max_{j}\left\{\widehat{\tau}_{nj}\right\}$ occurs in component $g$} \\
 0 & \text{otherwise.}
\end{array}
\right. 
\]
Then, the hierarchical initialization procedure proceeds according to the scheme in \figurename~\ref{fig:network}, where each arrow is directed from the model used for initialization to the model to be estimated.
Thus, $\widehat{z}_{ng}^{\text{$NN$-VE}}$ is used to initialize the EM of both $tN$-VE and $Nt$-VE, obtaining $\widehat{z}_{ng}^{\text{$tN$-VE}}$ and $\widehat{z}_{ng}^{\text{$Nt$-VE}}$, respectively, while $\widehat{z}_{ng}^{\text{$NN$-EV}}$ is used to initialize the EM of both $tN$-EV and $Nt$-EV, leading to $\widehat{z}_{ng}^{\text{$tN$-EV}}$ and $\widehat{z}_{ng}^{\text{$Nt$-EV}}$, respectively.
Also, following the same principle, the model between $NN$-VE and $NN$-EV leading to the maximum likelihood is used to initialize the EM for $NN$-VV.
Without going into further details on this hierarchical procedure, in the last step the model between $Nt$-VV, $tN$-VV, $tt$-VE, and $tt$-EV leading to the maximum likelihood is used to initialize the EM of $tt$-VV.

\subsection{Convergence criterion}
\label{subsec:Convergence Criterion}

The Aitken acceleration procedure \citep{Aitk:OnBe:1926} is used to estimate the asymptotic maximum of the log-likelihood at each iteration of the EM algorithm. 
Based on this estimate, a decision can be made regarding whether or not the algorithm has reached convergence;
that is, whether or not the log-likelihood is sufficiently close to its estimated asymptotic value. 
The Aitken acceleration at iteration $k$ is given by
\begin{displaymath}
	a^{\left(k\right)}=\frac{l^{\left(k+1\right)}-l^{\left(k\right)}}{l^{\left(k\right)}-l^{\left(k-1\right)}},
\end{displaymath}
where $l^{\left(k+1\right)}$, $l^{\left(k\right)}$, and $l^{\left(k-1\right)}$ are the log-likelihood values from iterations $k+1$, $k$, and $k-1$, respectively. 
Then, the asymptotic estimate of the log-likelihood at iteration $k + 1$ \citep{Bohn:Diet:Scha:Schl:Lind:TheD:1994} is given by
\begin{displaymath}	l_{\infty}^{\left(k+1\right)}=l^{\left(k\right)}+\frac{1}{1-a^{\left(k\right)}}\left(l^{\left(k+1\right)}-l^{\left(k\right)}\right).
\end{displaymath}
In the analyses in Section~\ref{sec:real datas}, we follow \citet{McNi:Mode:2010} and stop our algorithms when $l_{\infty}^{\left(k+1\right)}-l^{\left(k\right)}<\epsilon$, with $\epsilon=0.05$.

\section{Model selection and clustering performance}
\label{sec:model selection}

In model-based clustering, model selection criteria are commonly used to choice the best model and to select the number of groups.
Among them,  we will adopt the Bayesian information criterion \citep[BIC;][]{Schw:1978}
\begin{displaymath}
\text{BIC}= 2 l\left(\textsubtilde{$\widehat{\bpsi}$}\right) - m \ln N,
\end{displaymath}
where \textsubtilde{$\widehat{\bpsi}$} is the ML estimate of $\textsubtilde{$\bpsi$}$, $l\left(\textsubtilde{$\widehat{\bpsi}$}\right)$ is the maximized observed-data log-likelihood, and $m$ is the overall number of free parameters in the model (see the last three columns in \tablename~\ref{tab:parsimonious models}), and the integrated completed likelihood \citep[ICL;][]{Bier:Cele:Gova:2000} in the formulation given by \citet{Andr:McNi:Exte:2011}
\begin{equation}
\text{ICL}\approx \text{BIC} + \sum_{n=1}^{N} \sum^{G}_{g=1} \text{MAP}\left(\widehat{\tau}_{ng}\right) \ln\widehat{\tau}_{ng}.
\label{eq:ICL}
\end{equation}
A different ICL definition is used by \citet{Baek:McLa:Mixt:2011}.
The two definitions differ on whether or not it is the MAP of the fuzzy clustering in the first part of the entropy.
It is not immediately clear from \citet{Bier:Cele:Gova:2000} which definition is correct.
We have chosen the formulation in \eqref{eq:ICL} because it appears more widely adopted in literature \citep[see, e.g.,][]{McNi:Murp:Pars:2008,McNi:Murp:Mode:2010,McNi:Sube:Clus:2012}. 

In order to evaluate the clustering performance in cases in which the true classification is known, the adjusted Rand index \citep[ARI;][]{Hube:Arab:Comp:1985}, and the  misclassification rate  will be taken into account.
We recall that the ARI has an expected value of 0 and perfect classification would result in a value equal to 1.

\section{Applications to real data}
\label{sec:real datas}

This section illustrates some real data applications of the family of linear CWMs defined in Section~\ref{sec:The parsimonious Student-t LCWMs}.

\subsection{Student data}
\label{subsec:Student Data}

The first application concerns data coming from a survey of $N=270$ students attending a statistics course at the Department of Economics and Business of the University of Catania in the academic year 2011/2012.
The questionnaire included seven items, but the analysis we present below only concerns the following subset of variables:
\begin{align*}
\text{GENDER}&=\text{gender of the respondent;}\\
\text{HEIGHT}&=\text{height of the respondent, measured in centimeters;}\\
\text{WEIGHT}&=\text{weight of the respondent, measured in kilograms;}\\
\text{HEIGHT.F}&=\text{height of respondent's father, measured in centimeters.}
\end{align*}
There are $G=2$ groups of respondents with respect to the GENDER variable: $N_M=119$ males and $N_F=151$ females.
The considered data are available at \url{http://www.economia.unict.it/punzo/}. 
In the following, the two groups will be simply referred to as $G_M$ and $G_F$, respectively.
Moreover, we shall focus first on the joint distributions of WEIGHT and HEIGHT, then on HEIGHT and HEIGHT.F.
In both scenarios, data will be assumed unlabeled with respect to GENDER.
However, the true labels will be useful for evaluating the quality of the obtained clustering.

\subsubsection{First scenario: HEIGHT and WEIGHT} 
\label{subsubsec:first scenario: HEIGHT and WEIGHT}

\begin{figure}[!ht]
\centering
\resizebox{0.76\textwidth}{!}{
\includegraphics{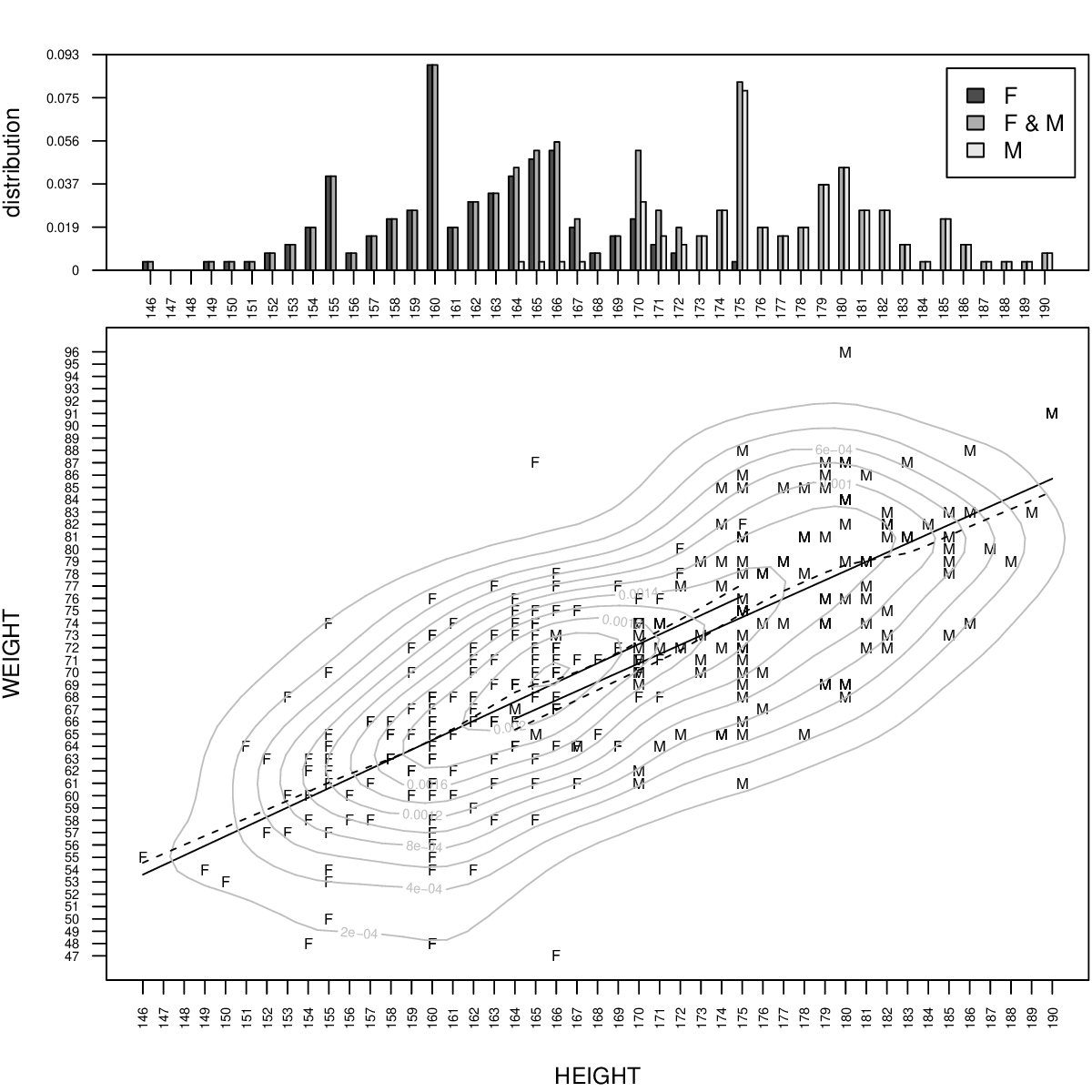} 
}
\caption{
Student Data: CW-plot of HEIGHT and WEIGHT for $119$ male, and $151$ female, students ($\mathsf{M}$ denotes male and $\mathsf{F}$ female).
}
\label{fig:WEIGHTandHEIGHT}
\end{figure}

\figurename~\ref{fig:WEIGHTandHEIGHT} concerns the observed labeled data.
This graphical representation will be simply referred to as the CW-plot.
The top of \figurename~\ref{fig:WEIGHTandHEIGHT} displays a bar plot of the HEIGHT variable, including the overall empirical marginal density as well as the empirical marginal densities, for $G_M$ and $G_F$, weighted according to their sizes; bars are color-coded, using a gray scale, with respect to the GENDER variable.
We remark that many students tend to approximate their height to ``classical'' values, such as 155, 160, 170, 175, and so on.
For classification purposes, the variable HEIGHT separates the two groups quite well.
The bottom of \figurename~\ref{fig:WEIGHTandHEIGHT} is a scatter plot of HEIGHT and WEIGHT, where male and female students are labeled with $\mathsf{M}$ and $\mathsf{F}$, respectively.
We give the isodensities of a bivariate normal kernel estimator as computed by the function \texttt{bkde2D} of the R-package \texttt{KernSmooth} \citep[see, e.g.,][]{Wand:Jone:Kern:1995}.
The plot also shows the functional dependence of WEIGHT on HEIGHT separately for $G_M$ and $G_F$; the solid lines concern the linear regression models while the dashed ones arise from a locally-weighted polynomial regression computed using the \texttt{lowess} function of the R-package \texttt{stats} \citep[see][for details]{Clev:Robu:1979}.
A simple visual comparison between solid and dashed lines justifies the linearity assumption of WEIGHT on HEIGHT, underlying the linear CWMs of the proposed family.
Moreover, the regression lines in \figurename~\ref{fig:WEIGHTandHEIGHT} seem to indicate that these models have the same parameters in $G_M$ and $G_F$.
In these terms note also that:
\begin{enumerate}
	\item the $t$-test for equal slopes provides a $p$-value of 0.147,
	\item the $t$-test for equal intercepts provides a $p$-value of 0.364, and
	\item the F-test of homoscedasticity of residuals in the two groups provides a $p$-value of 0.992.
\end{enumerate}

Now, let us ignore the true classification induced by GENDER and fit the data according to the linear CWMs in \tablename~\ref{tab:parsimonious models} by using the true value $G=2$.
\tablename~\ref{tab:scenario 1 - IC criteria} lists the values of the BIC, ICL, and ARI for the twelve models.
\begin{table}[!ht]
\caption{Student Data: Values of the BIC, ICL, and ARI ($G=2$).
Bold numbers highlight the best model for each criterion/index.
\label{tab:scenario 1 - IC criteria}
}
\centering
\subtable[BIC]{
\label{tab:scenario 1 - BIC}
\resizebox{!}{0.06\textheight}{
\begin{tabular}{cccc}
\toprule
    &	VE	&	EV	&	VV	\\
\midrule															
$NN$	&	\textbf{-3726.197}	&	-3756.561	&	-3742.947	\\
$tN$	&	-3737.394	&	-3762.160	&	-3754.144	\\
$Nt$	&	-3731.795	&	-3766.517	&	-3749.642	\\
$tt$	&	-3742.992	&	-3772.115	&	-3760.839  \\
\bottomrule	
\end{tabular}
}
}
\subtable[ICL]{\label{tab:scenario 1 - ICL}
\resizebox{!}{0.06\textheight}{
\begin{tabular}{cccc}
\toprule
    &	VE	&	EV	&	VV	\\
\midrule
$NN$	&	\textbf{-3750.466}	&	-3880.260	&	-3767.213	\\
$tN$	&	-3761.663	&	-3885.858	&	-3778.409	\\
$Nt$	&	-3756.064	&	-3869.845	&	-3773.484	\\
$tt$	&	-3767.261	&	-3875.443	&	-3784.681	\\
\bottomrule	
\end{tabular}
}
}
\subtable[ARI]{\label{tab:scenario 1 - ARI}
\resizebox{!}{0.06\textheight}{
\begin{tabular}{cccc}
\toprule
    &	VE	&	EV	&	VV	\\
\midrule															
$NN$	&	0.750	&	0.008	&	0.750	\\
$tN$	&	0.750	&	0.008	&	0.750	\\
$Nt$	&	0.750	&	0.005	&	\textbf{0.776}	\\
$tt$	&	0.750	&	0.005	&	\textbf{0.776}	\\
\bottomrule	
\end{tabular}
}
}
\end{table}

\begin{figure}[!ht]
\centering
\resizebox{0.76\textwidth}{!}{
\includegraphics{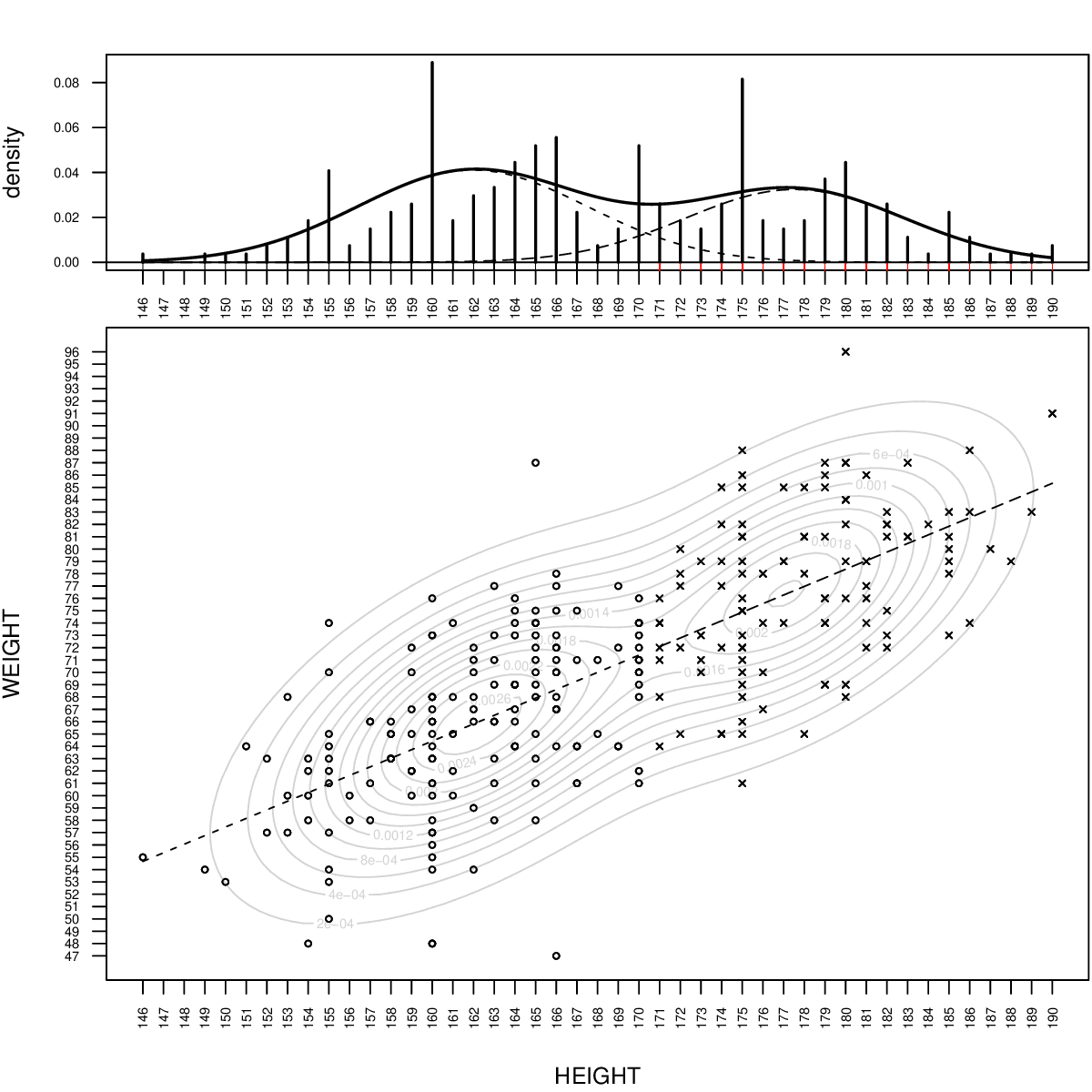} 
}
\caption{
Student Data: CW-plot of HEIGHT and WEIGHT for $NN$-VE ($G=2$).
}
\label{fig:scenario1-NN-VE}
\end{figure}

\begin{figure}[!ht]
\centering
\resizebox{0.7\textwidth}{!}{
\includegraphics{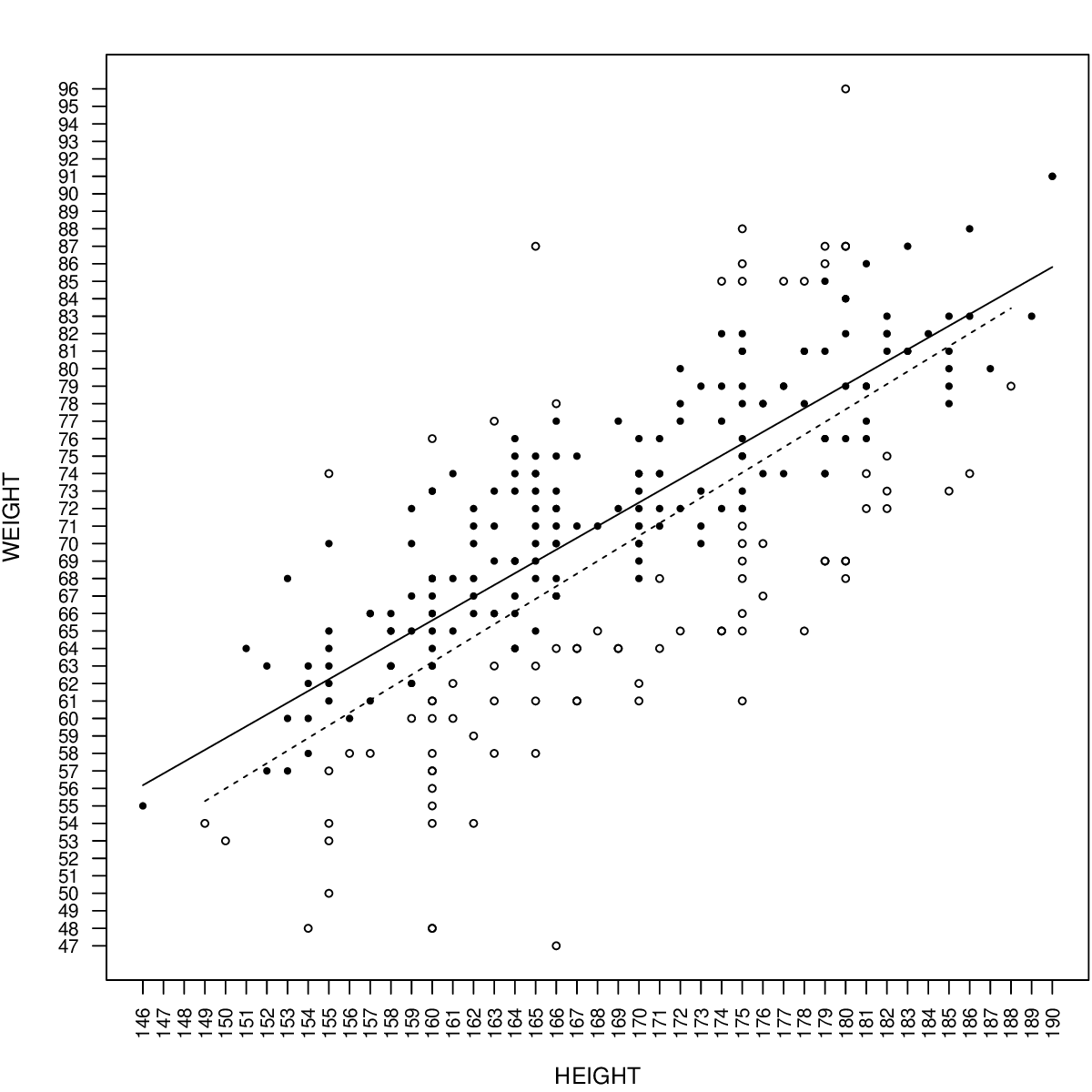} 
}
\caption{
Student Data: Scatter plot of WEIGHT versus HEIGHT.
The two types of lines and symbols displayed arise from the fit of a mixture of $G=2$ Gaussian regressions.
}
\label{fig:WH_FMR}
\end{figure}

$NN$-VE (Gaussian marginal and conditional component densities and equal linear model between clusters) is the best model according to both BIC (-3726.197) and ICL (-3750.466).
The corresponding CW-plot is displayed in \figurename~\ref{fig:scenario1-NN-VE}. 
As for the ARI is concerned, in practice we have similar results for all models of type VE and VV. 
Thus,  the group structure of the data is due to different intra-group distributions for the covariates, while the linear relationship is homogenous.
In other words, this is a case of assignment dependence that a standard finite mixture of linear regressions is not able to represent.
In order to show it empirically, we have also fitted a mixture of $G=2$ linear Gaussian regressions by means of the \texttt{flexmix} function of the R-package \texttt{flexmix} \citep{Leis:Flex:2004}.
The group-conditional distribution of $Y|X$ is  Gaussian like in $NN$-VE.
\figurename~\ref{fig:WH_FMR} highlights that the mixture model with a fixed covariate is not able to recognize the group-structure of the data. 
This is also confirmed by an ARI value equal to 0.00288.

\subsubsection{Second scenario: HEIGHT.F and HEIGHT} 
\label{subsubsec:second scenario: HEIGHT.F and HEIGHT}

\begin{figure}[!ht]
\centering
\resizebox{0.76\textwidth}{!}{
\includegraphics{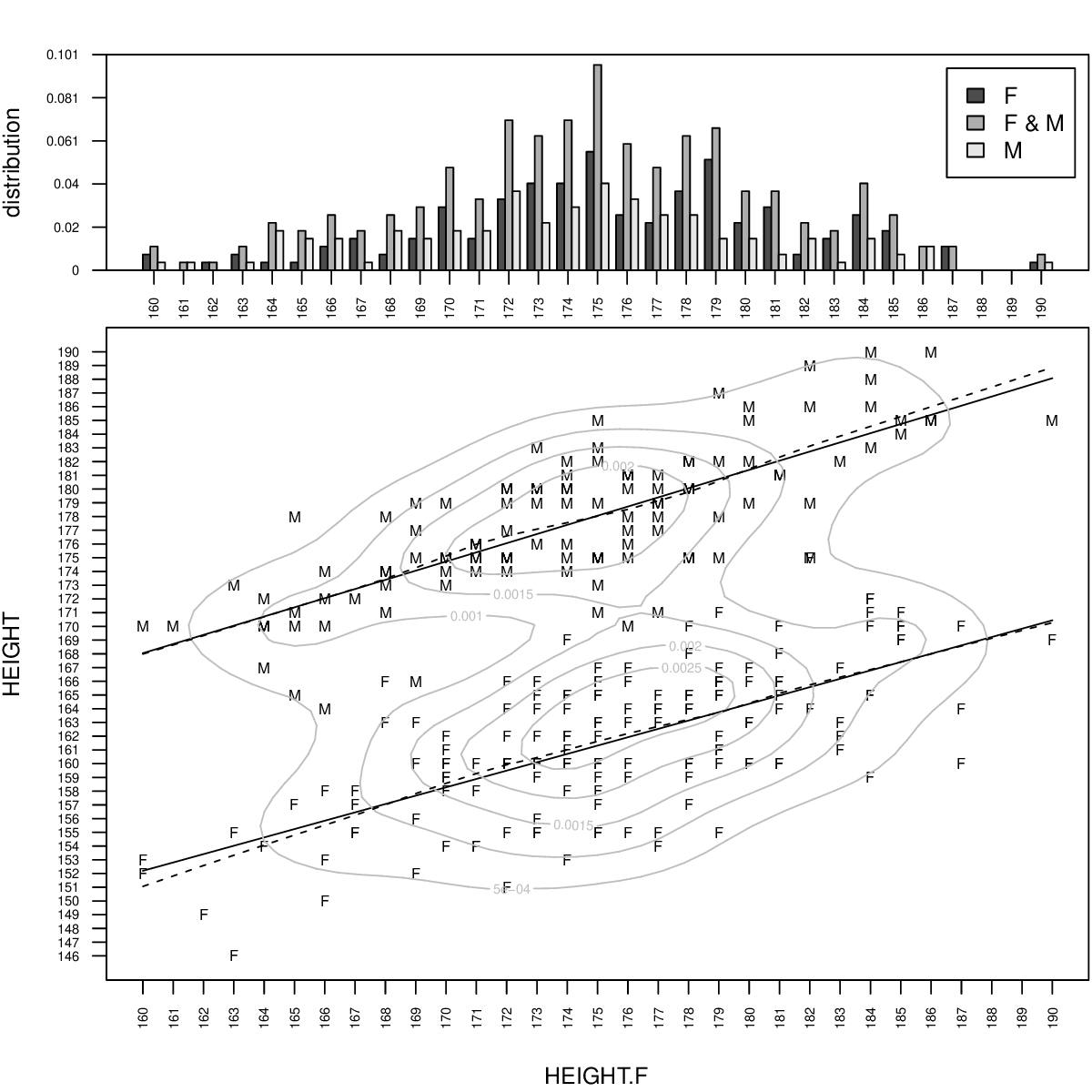} 
}
\caption{
Student Data: CW-plot of HEIGHT and HEIGHT.F for $119$ male and $151$ female, students ($\mathsf{M}$ denotes male and $\mathsf{F}$ female).
}
\label{fig:HEIGHT.F and HEIGHT}
\end{figure}

\figurename~\ref{fig:HEIGHT.F and HEIGHT} shows the CW-plot of HEIGHT.F and HEIGHT by considering the classification induced by GENDER.
Although, also in this case, linearity between variables appear to be reasonable, the linear models for the two groups differ, especially in terms of intercept. 
Note also that, the $F$-test of homoscedasticity of the residuals in the two groups gives a $p$-value of 0.086 while the $t$-tests for equal slopes and equal intercepts provide practically null $p$-values.

\begin{figure}[!ht]
\centering
\resizebox{0.76\textwidth}{!}{
\includegraphics{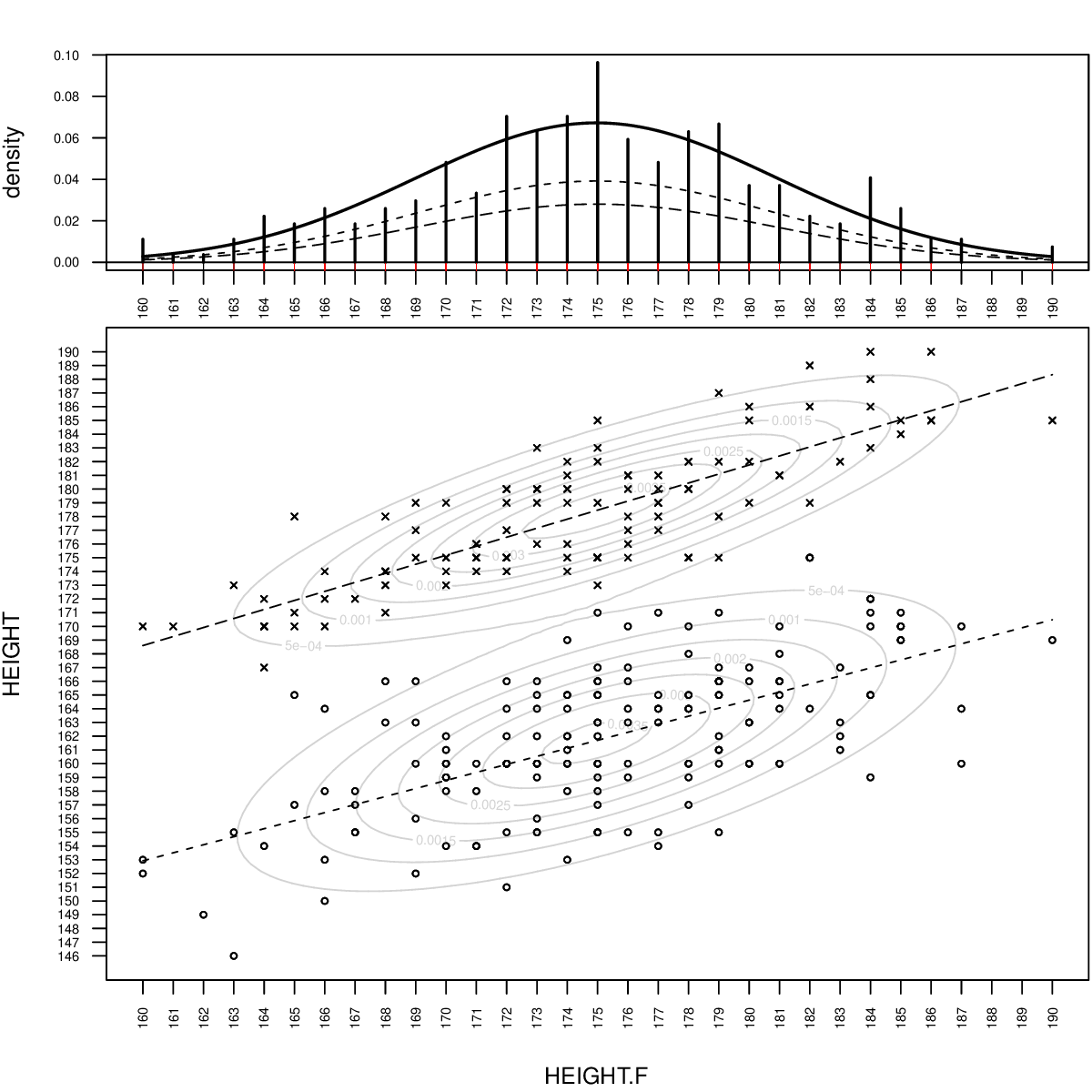} 
}
\caption{
Student Data:  CW-plot of HEIGHT.F and HEIGHT for NN-EV ($G=2$).
}
\label{fig:scenario2-NN-EV}
\end{figure}

As in Section~\ref{subsubsec:first scenario: HEIGHT and WEIGHT}, we fit the linear CWMs, with $G=2$, ignoring the true classification induced by GENDER. 
The values of BIC, ICL, and ARI for the twelve models are given in
\tablename~\ref{tab:scenario 2 - IC criteria}.
\begin{table}[!ht]
\caption{Student Data: Values of the BIC, ICL, and ARI ($G=2$).
Bold numbers highlight the best model for each criterion/index.
\label{tab:scenario 2 - IC criteria}
}
\centering
\subtable[BIC]{
\label{tab:scenario 2 - BIC}
\resizebox{!}{0.06\textheight}{
\begin{tabular}{cccc}
\toprule
    &	VE	&	EV	&	VV	\\
\midrule															
$NN$	&	-3726.339	&	\textbf{-3594.401}	&	-3601.955	\\
$tN$	&	-3737.536	&	-3599.999	&	-3613.152	\\
$Nt$	&	-3731.937	&	-3605.598	&	-3613.152	\\
$tt$	&	-3743.134	&	-3611.196	&	-3624.348	\\
\bottomrule	
\end{tabular}
}
}
\subtable[ICL]{
\label{tab:scenario 2 - ICL}
\resizebox{!}{0.06\textheight}{
\begin{tabular}{cccc}
\toprule
    &	VE	&	EV	&	VV	\\
\midrule
$NN$	&	-3822.623	&	\textbf{-3597.252}	&	-3605.016\\
$tN$	&	-3833.820	&	-3602.850	&	-3616.212\\
$Nt$	&	-3828.221	&	-3608.449	&	-3616.212\\
$tt$	&	-3839.418	&	-3614.047	&	-3627.409\\
\bottomrule	
\end{tabular}
}
}
\subtable[ARI]{
\label{tab:scenario 2 - ARI}
\resizebox{!}{0.06\textheight}{
\begin{tabular}{cccc}
\toprule
    &	VE	&	EV	&	VV	\\
\midrule															
$NN$	&	0.009	&	0.898	&	\textbf{0.912}	\\
$tN$	&	0.009	&	0.898	&	\textbf{0.912}	\\
$Nt$	&	0.009	&	0.898	&	\textbf{0.912}	\\
$tt$	&	0.009	&	0.898	&	\textbf{0.912}	\\
\bottomrule	
\end{tabular}
}
}
\end{table}
In this case, the best model is $NN$-EV (see also the corresponding CW-plot in \figurename~\ref{fig:scenario2-NN-EV}).
The fitted model also appears to be a good compromise in terms of the ARI values of \tablename~\ref{tab:scenario 2 - ARI}.
Differently from the first scenario, here the group-structure is due to the different intra-group linear models, while the distribution of the covariate is homogenous. 
This is an example of assignment independence which can be recognized by a simple mixture of $G=2$ linear (Gaussian) regressions too.

\subsection{Tourist data}
\label{subsec:tourism data}

\begin{figure}[!ht]
\centering
\resizebox{0.76\textwidth}{!}{
\includegraphics{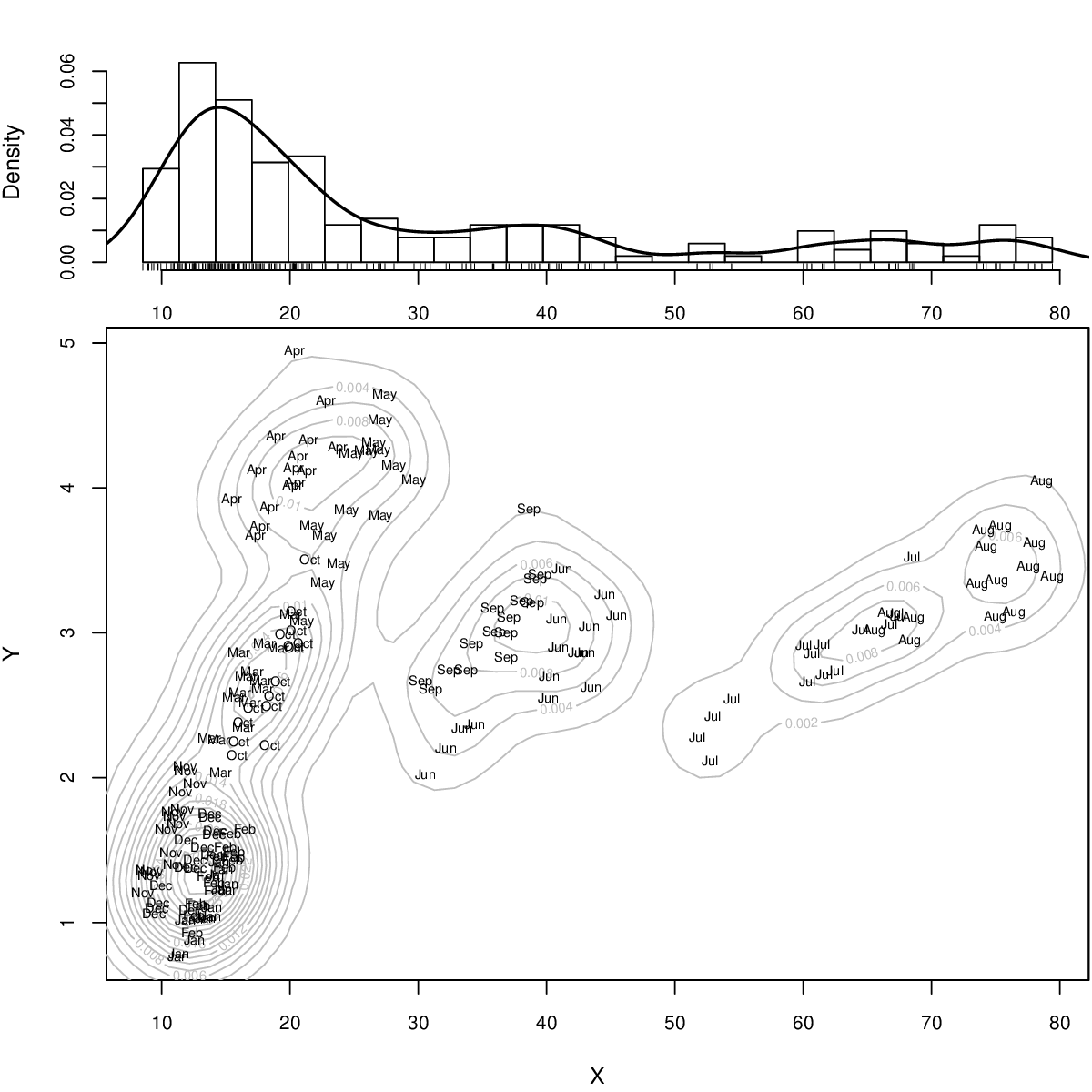} 
}
\caption{
Tourist data: CW-plot of \textit{tourist overnights} ($X$, in millions) and \textit{attendance at museums and monuments} ($Y$, in millions) in Italy over the period from January 1996 to December 2010 ($N=180$).
The univariate normal kernel density of $X$ is superimposed on the histogram.
The isodensities from a bivariate normal kernel density estimator are also visualized on the scatter plot.
Month abbreviations are used as labels in the scatter plot. 
}
\label{fig:tourism.real.data}
\end{figure}

\begin{figure}[!ht]
\centering
\subfigure[BIC\label{fig:tourism.BIC}]
{\includegraphics[width=0.49\textwidth]{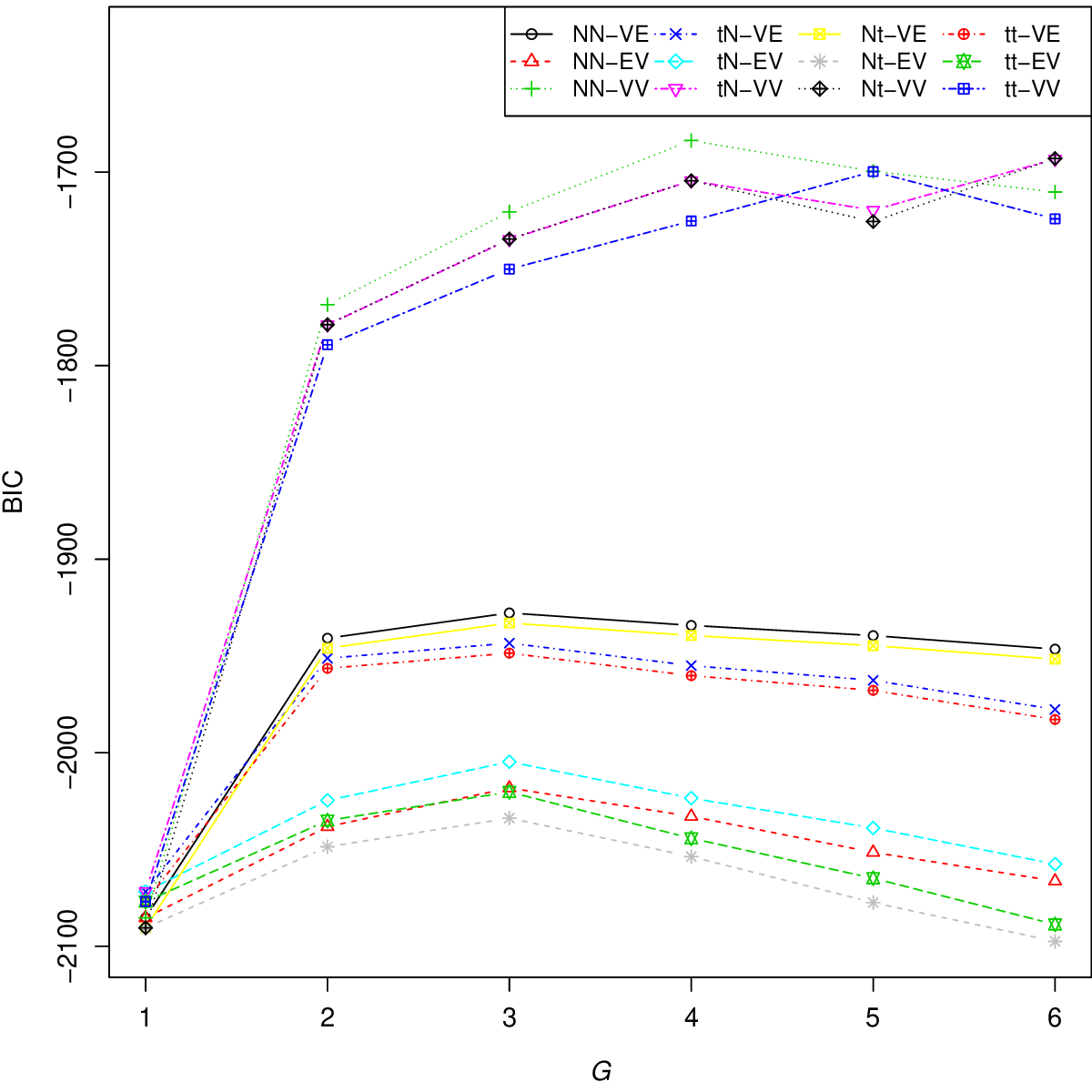}}
\subfigure[ICL\label{fig:tourism.ICL}]
{\includegraphics[width=0.49\textwidth]{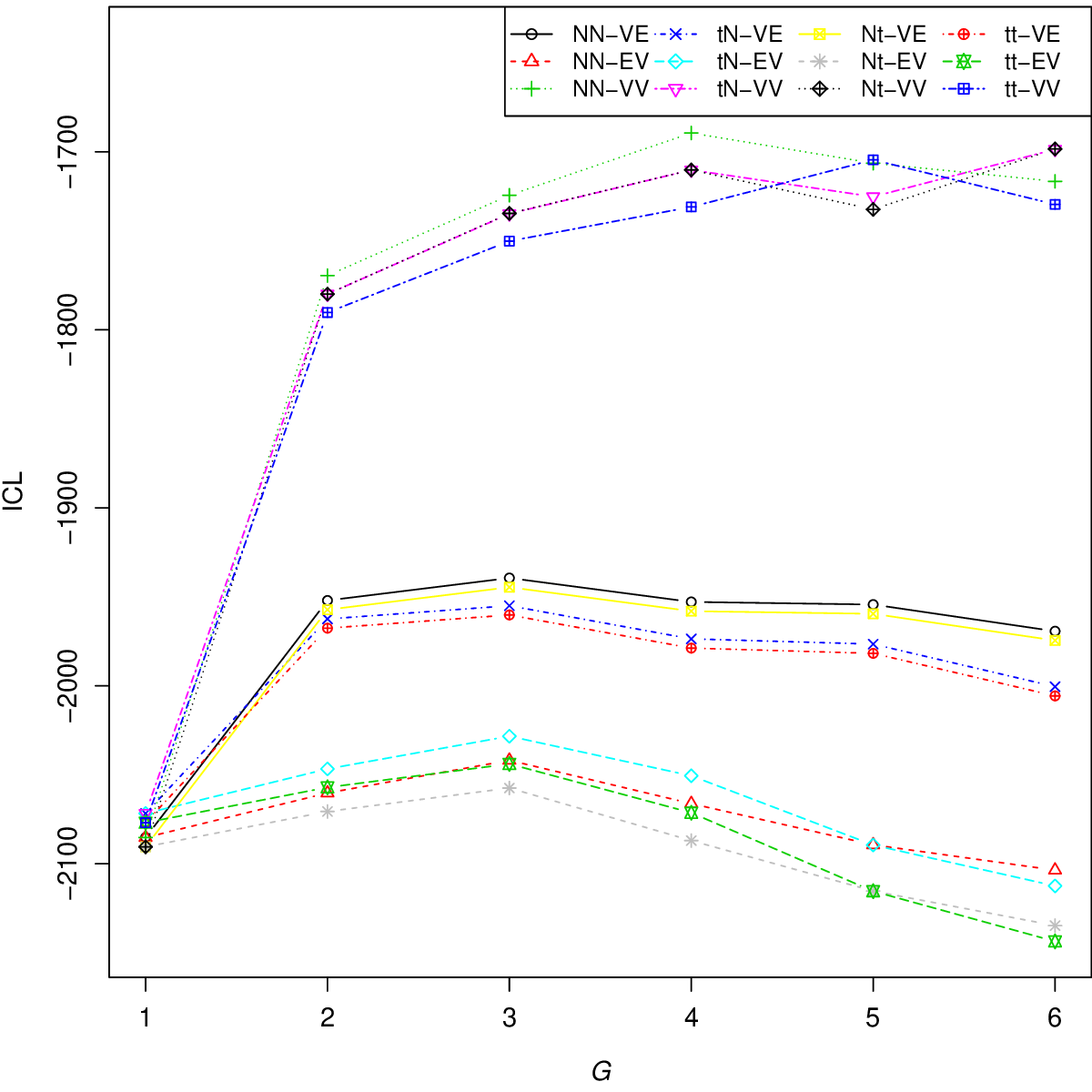}}
\caption{
Tourist data: Values of the BIC and ICL ($G=1,\ldots,6$).
}
\label{fig:tourism IC}
\end{figure}

\begin{figure}[!ht]
\centering
\resizebox{0.76\textwidth}{!}{
\includegraphics{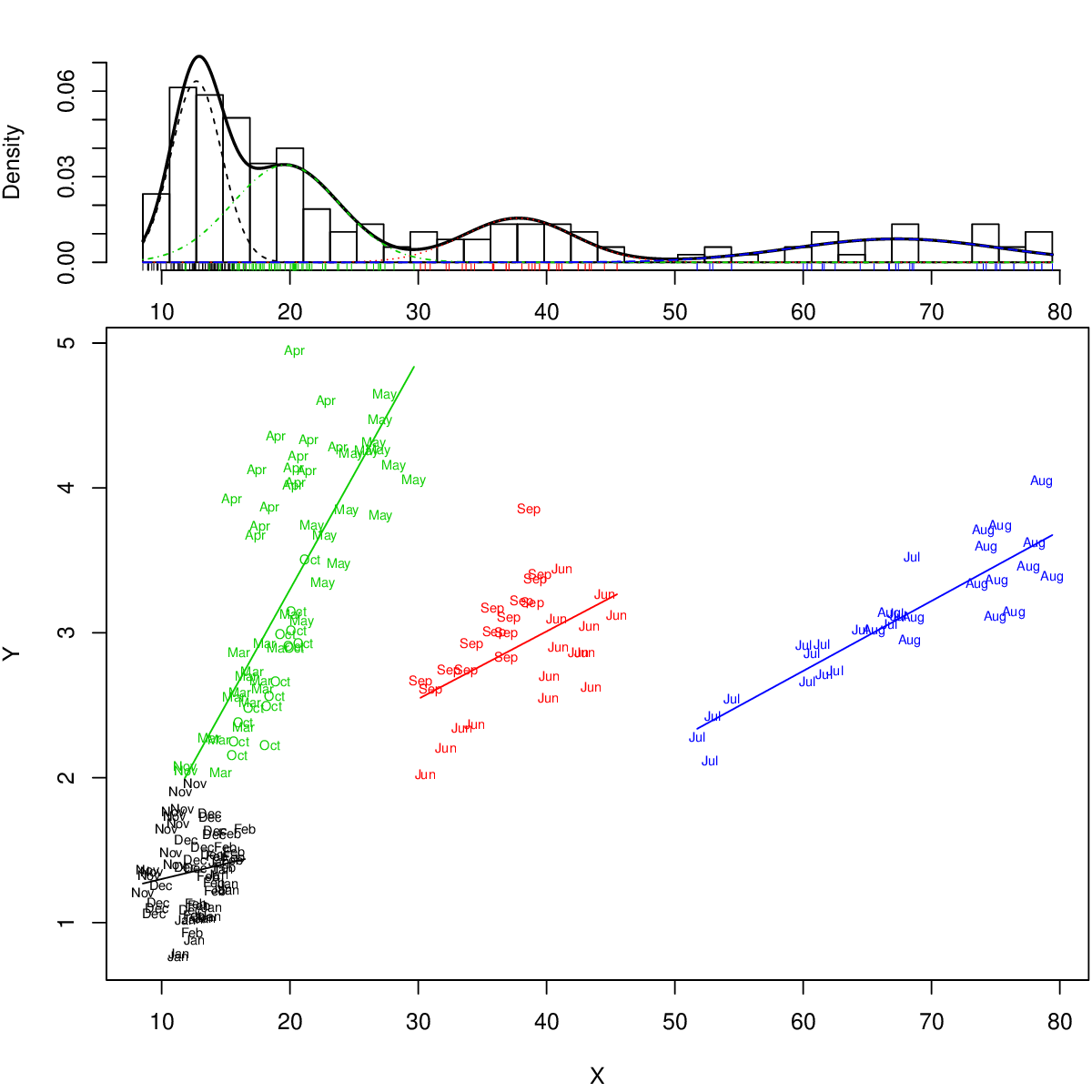} 
}
\caption{
Tourist data: CW-plot of model $NN$-VV with $G=4$ components  ($X=$ ``tourist overnights'', in millions, and $Y=$ ``attendance at museums and monuments'', in millions).
}
\label{fig:tourism.CWM}
\end{figure}

The second application focuses on $N=180$ monthly data (tourism data) concerning \textit{tourist overnights} ($X$, data in millions) and \textit{attendance at museums and monuments} ($Y$, data in millions) in Italy over the 15-year period spanning from January 1996 to December 2010.
These data have been recently analyzed by \citet{Cell:Cucc:tour:2013} and are available at \url{http://www.robertocellini.it/doc/master_specializzazione/Cellini-Cuccia_ApEc2013_data1996-2010.pdf
}. 
The CW-plot of the labeled data (with respect to months) is shown in \figurename~\ref{fig:tourism.real.data}. 
It is straightforward to note how the heterogeneity of the data reveals a clear group-structure.
\figurename~\ref{fig:tourism IC} shows the values of the BIC and the ICL for the models in the proposed family of linear CWMs with $G$ ranging from 1 to 6.
Both criteria (BIC=-1683.727 and ICL=-1689.386) suggest the $NN$-VV, with $G=4$ components, displayed in \figurename~\ref{fig:tourism.CWM}. 
Here, it is interesting to analyze the relationship between the obtained clusters -- characterized by 4 different slopes -- and the time-covariate (months; see \tablename~\ref{fig:tourism table}).
\begin{table}[!ht] 
\centering
\begin{tabular}{ccrrrrrrrrrrrr}
  \toprule
group  &&  Jan & Feb & Mar & Apr & May & Jun & Jul & Aug & Sep & Oct & Nov & Dec \\ 
 \midrule
1 &&  15 &  15 &   0 &   0 &   0 &   0 &   0 &   0 &   0 &   0 &  13 &  15 \\ 
2 &&   0 &   0 &   0 &   0 &   0 &  15 &   0 &   0 &  15 &   0 &   0 &   0 \\ 
3 &&   0 &   0 &  15 &  15 &  15 &   0 &   0 &   0 &   0 &  15 &   2 &   0 \\ 
4 &&   0 &   0 &   0 &   0 &   0 &   0 &  15 &  15 &   0 &   0 &   0 &   0 \\ 
\bottomrule
\end{tabular}
\caption{Tourist data: 
Relation between the $G=4$ clusters, obtained with the fitted $NN$-VV, and the variable time identified by month. 
}
\label{fig:tourism table}
\end{table}
The four clusters, arising from the $NN$-VV, are almost perfectly related to the months (except for two units in November, which concern years 2006 and 2010). 
In particular, we have: 
\begin{description}
\item[Group 1]: units from November to February, 
\item[Group 2]: units in June and September, 
\item[Group 3]: units in March, April, May, and October, and 
\item[Group 4]: units in July and August. 
\end{description}

This is an example in which the group structure of the data is due to differences both in the intra-group marginal distributions and the linear models.

\subsection{Crab data}
\label{subsec:Usefulness of the linear t CWM}

The third application, based on the very popular crab data set of \citet{Camp:Maho:amul:1974} on the genus \textit{Leptograpsus}, has the aim of showing that the $t$-based linear CWMs ($tN$-VE, $tN$-EV, $tN$-VV, $Nt$-VE, $Nt$-EV, $Nt$-VV, $tt$-VE, $tt$-EV, and $tt$-VV) can provide more robust classification than the linear (completely) Gaussian ones ($NN$-VE, $NN$-EV, and $NN$-VV).
Attention is focused on the sample of $N=100$ blue crabs, there being $N_1=50$ males (group 1) and $N_2=50$ females (group 2).
Each specimen having $p=2$ measurements (in millimeters): the rear width ($\text{RW}=Y$) and the length along the midline ($\text{CL}=X$) of the carapace. 

\begin{figure}[!ht]
\centering
\subfigure[$y_{25}=-15$\label{fig:C1}]
{\includegraphics[width=0.49\textwidth]{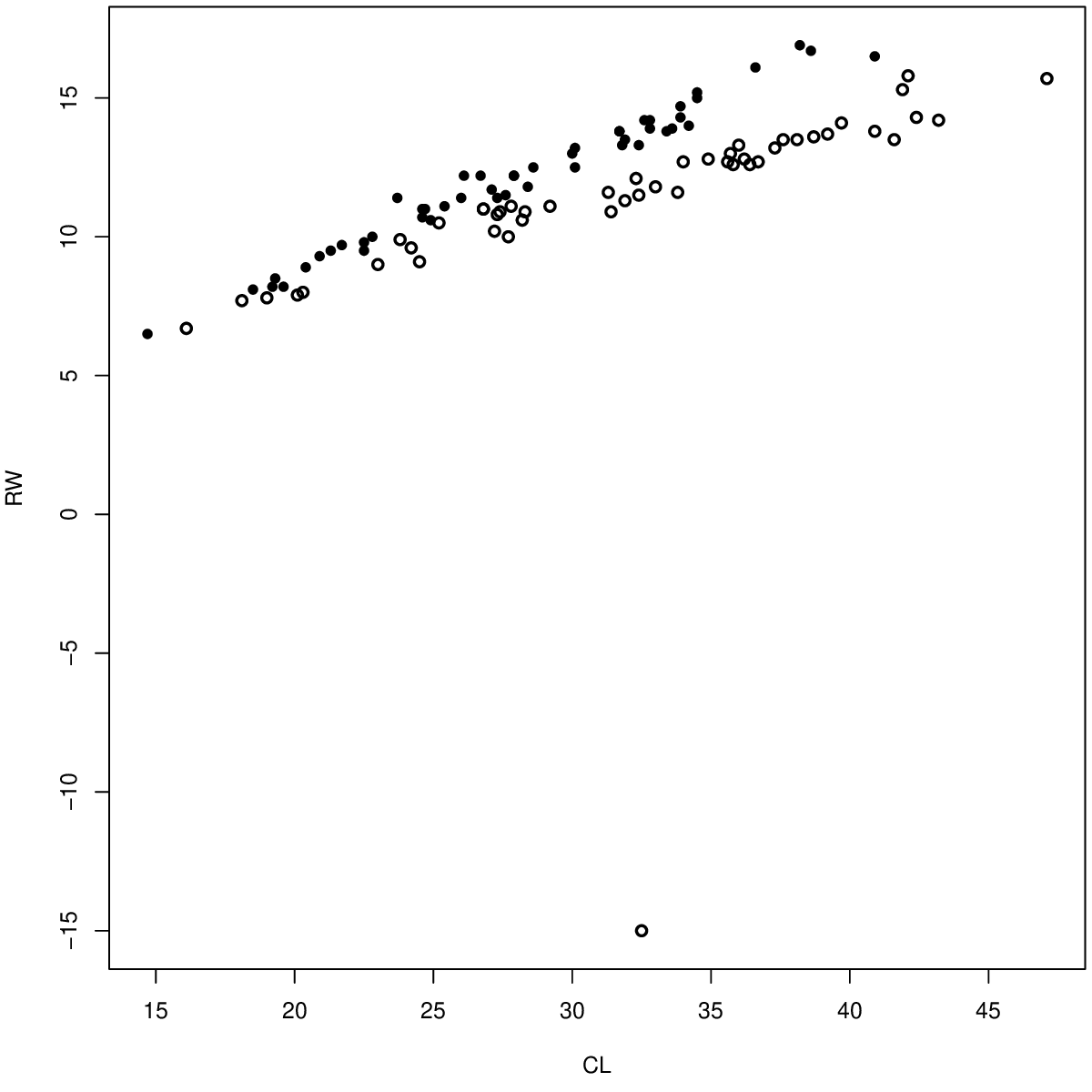}}
\subfigure[$y_{25}=-10$\label{fig:C2}]
{\includegraphics[width=0.49\textwidth]{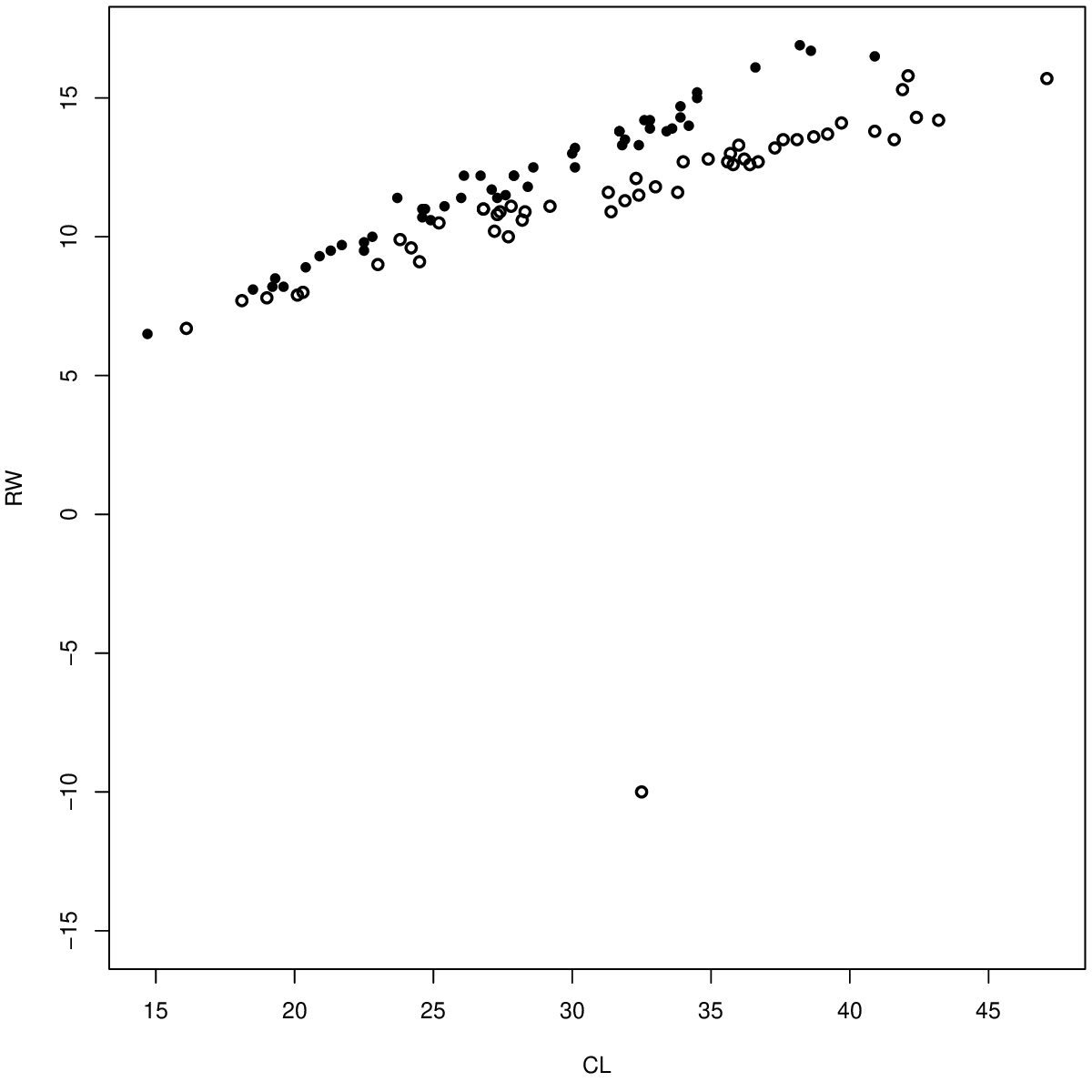}}
\subfigure[$y_{25}=-5$\label{fig:C3}]
{\includegraphics[width=0.49\textwidth]{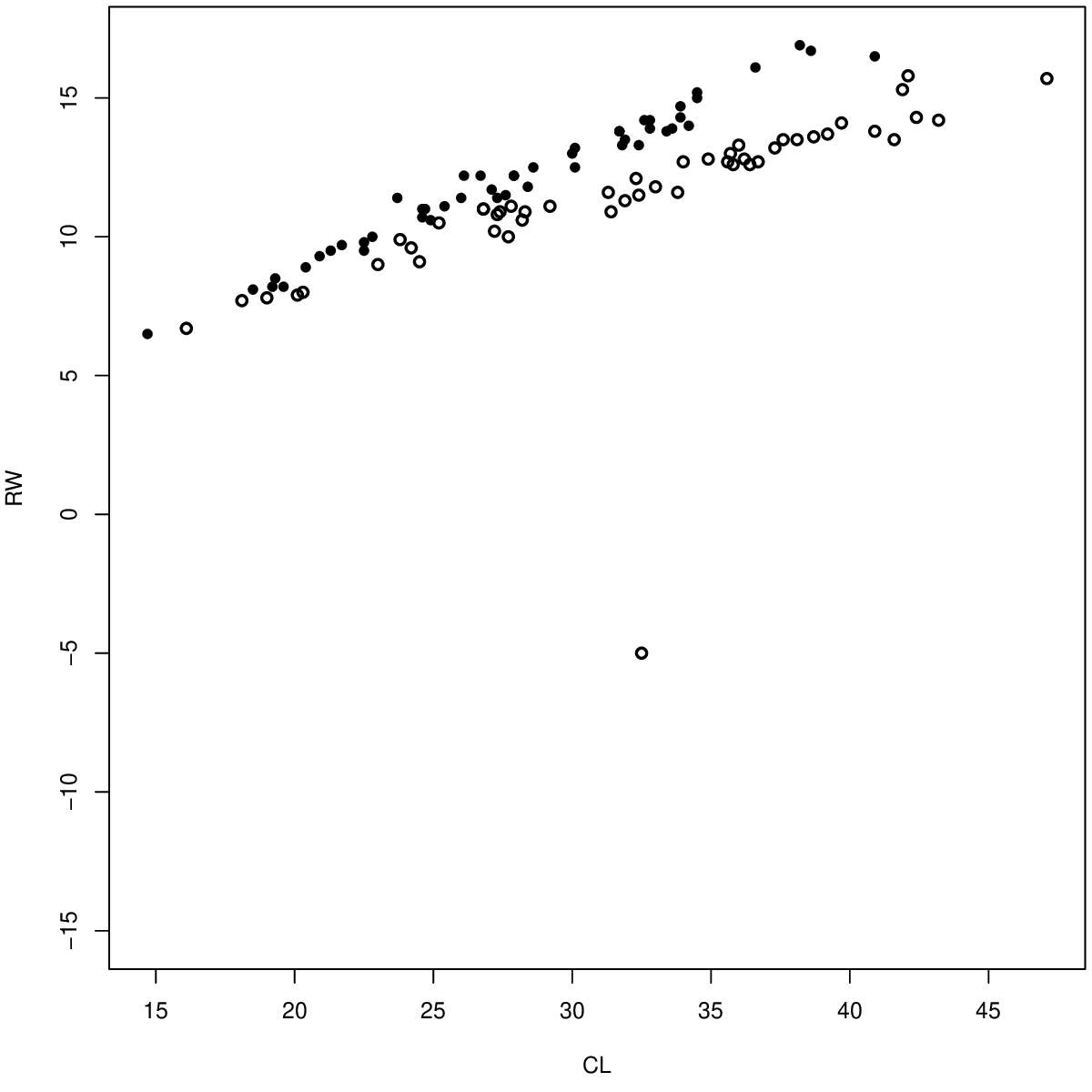}}
\subfigure[$y_{25}=0$\label{fig:C4}]
{\includegraphics[width=0.49\textwidth]{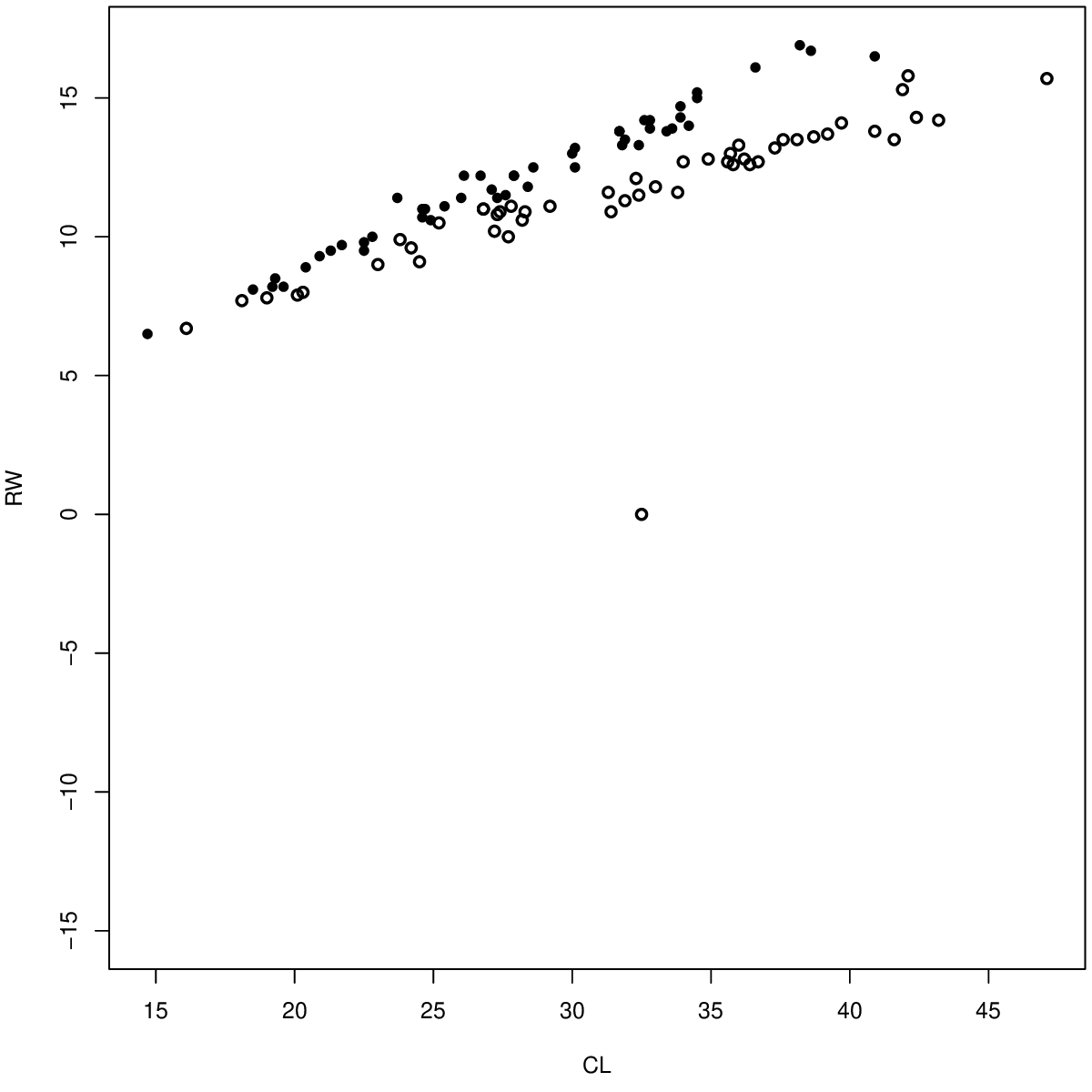}}
\caption{
Scatter plots of the sample of $N=100$ blue crabs with different values for $y_{25}$.
The variables are rear width (RW) and length along the midline (CL) of the carapace, for $N_1=50$ males and $N_2=50$ females ($\boldsymbol{\circ}$ denotes male and $\bullet$ female).
}
\label{fig:crab with outliers}
\end{figure}

Following the scheme of \citet[][Section~7.8]{McLa:Peel:fini:2000}, some outliers were introduced by substituting the original value of $y_{25}$ (11.9) with some atypical values (-15, -10, -5, and 0).
This leads to four different ``perturbed'' data sets which are displayed in \figurename~\ref{fig:crab with outliers}.

\tablename~\ref{fig:crab clustering} reports the number of misallocated observations for each of the twelwe models and each perturbed version of the original data set. 
\begin{table}[!ht] 
\centering
\resizebox{\textwidth}{!}{\begin{tabular}{rc cccc cccccccccc ccc}
\toprule
&&	\multicolumn{3}{c}{linear Gaussian CWMs (A)} && \multicolumn{9}{c}{$t$-based linear CWMs (B)} &&		&	\\
$y_{25}$ 	&&	$NN$-VE	&	$NN$-EV	&	$NN$-VV	&&	$tN$-VE	&	$tN$-EV	&	$tN$-VV	&	$Nt$-VE	&	$Nt$-EV	&	$Nt$-VV	&	$tt$-VE	&	$tt$-EV	&	$tt$-VV	&&		$\min\left(A\right)$	&	$\min\left(B\right)$\\
\midrule
-15	&&	40	&	49	&	49	&&	40	&	49	&	49	&	40	&	\textbf{16}	&	49	&	40	&	\textbf{16}	&	49	&&		40	&	\textbf{16}	\\
-10	&&	40	&	49	&	50	&&	40	&	49	&	50	&	40	&	\textbf{16}	&	25	&	40	&	\textbf{16}	&	25	&&		40	&	\textbf{16}	\\
 -5	&&	40	&	49	&	50	&&	40	&	49	&	50	&	40	&	\textbf{13}	&	24	&	40	&	\textbf{13}	&	24	&&		40	&	\textbf{13}	\\
  0	&&	40	&	49	&	50	&&	40	&	49	&	50	&	40	&	\textbf{13}	&	21	&	40	&	\textbf{13}	&	21	&&		40	&	\textbf{13}	\\
\bottomrule
\end{tabular}
}
\caption{Crab data: 
Comparison of the number of misallocated observations when fitting the family of linear CWMs on the sample of $N=100$ blue crabs.
Bold numbers highlight the best results for each perturbed data set.
}
\label{fig:crab clustering}
\end{table}
Estimates are obtained by directly using $G=2$.
The last two columns report the minimum number of misallocated observations computed over the linear Gaussian CWMs and the $t$-based linear CWMs, respectively.
From the bold numbers in \tablename~\ref{fig:crab clustering} follows that some of the $t$-based linear CWMs, that is $Nt$-EV and $tt$-EV, are systematically more robust than the linear Gaussian CWMs (see also the results for $Nt$-VV and $tt$-VV).
In particular, since the perturbations are inserted ``vertically'' on the $Y$-variable, the best performers have the $t$ distribution for $p\left(y|x,\Omega_g\right)$, $g=1,2$.

\subsection{f.voles data}
\label{subsec:real data}

\begin{figure}[!ht]
	\vspace{-4mm}
	\begin{center}		
\resizebox{0.9\textwidth}{!}{\includegraphics{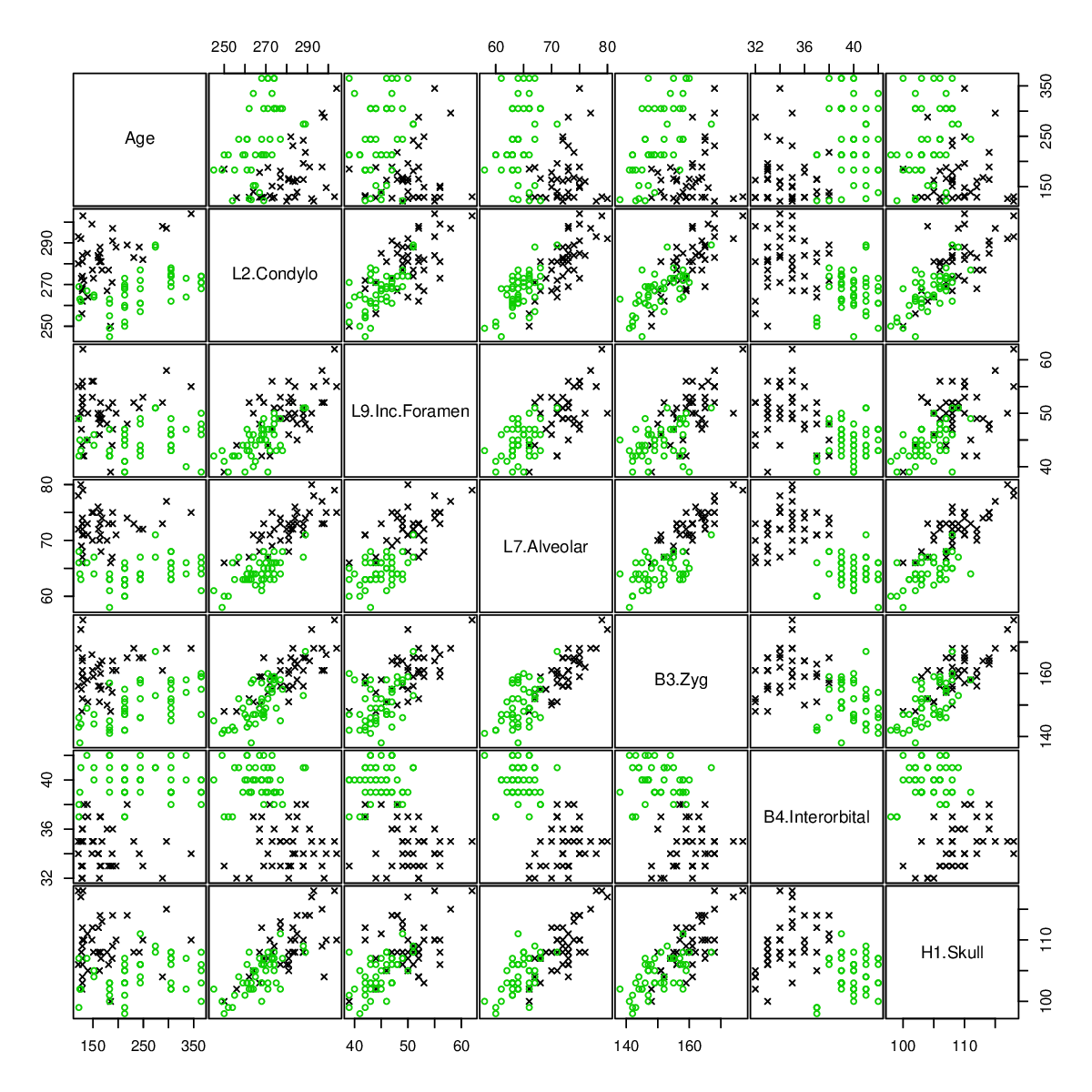}}
	\end{center}
	\vspace{-8mm}
	\caption{
	Scatter plot matrix of f.voles data	(\textcolor{green}{$\boldsymbol{\circ}$} and $\times$ denote species \textit{Microtus~ochrogaster} and \textit{M.~californicus}, respectively).
	}
	\label{fig:f.voles}
\end{figure}

The fourth application is based on the \texttt{f.voles} data set described in \citet*[][\tablename~5.3.7]{Flur:Afir:1997} and available in the R-package \texttt{Flury}.
This is an example with more than one covariate.
Data refer to measurements on $N=86$ female voles from two species, \textit{M.~californicus} ($N_1=45$) and \textit{M.~ochrogaster} ($N_2=45$). 
Variables  used here are: $\mathsf{Species}$ denoting the two species, $\mathsf{Age}$ measured in days, along with other six measurements related to skull (in units of 0.1 mm). The latter are named as in  \citet*{Airo:Hoff:Acom:1984}: $\mathsf{L}_2=\text{condylo-incisive length}$, $\mathsf{L}_9=\text{length of incisive foramen}$, $\mathsf{L}_7=\text{alveolar length of upper molar tooth row}$, $\mathsf{B}_3=\text{zygomatic width}$, $\mathsf{B}_4=\text{interorbital width}$, and $\mathsf{H}_1=\text{skull height}$. 
The  scatter plot matrix  for grouped-data is shown in \figurename~\ref{fig:f.voles}.

The purpose of \citet*{Airo:Hoff:Acom:1984} was to study age variability in \textit{M.~californicus} and \textit{M.~ochrogaster} and  predict age on the basis of the skull measurements. 
In this study, we assume that data are unlabelled with respect to $\mathsf{Species}$ and compare the classification provided by the three approaches: the family of linear CWMs, 
mixtures of linear Gaussian regressions (estimated by the R-package \texttt{flexmix}), and parsimonious mixtures of Gaussian distributions (estimated using the  R-package \texttt{mclust}; \citealp[see][for details]{Fral:Raft:Murp:Scru:mclu:2012}). 
For the first two classes of models, $\mathsf{Age}$ is the  response variable $Y$ and the $d=6$ skull measurements are the $\bX$ variable.
For parsimonious mixtures of Gaussian distributions, the vector $\left(Y,\bX'\right)'$ is considered as a whole. 
All the considered models have been fitted with $G=2$.

In the family of linear CWMs, the two models providing the largest values for the BIC and the ICL were $NN$-EV and $NN$-VE (BIC: $NN$-EV = $-3890.397$,
$NN$-VE = $-3895.917$; ICL: $NN$-VE = $-3896.143$, $NN$-EV = $-3902.788$). 
In particular, the two criteria selected a different model, although both the BIC and the ICL yielded quite close values for $NN$-EV and $NN$-VE.
On the contrary, the resulting  misclassification errors were very different:  $NN$-VE (selected by the ICL) yielded 
a perfect classification, while $NN$-EV (selected by the BIC) yielded a misclassification error of 38.37\%. 
A closer look to the membership probabilities showed that $NN$-VE led to a  sharp classification (the  entropy term in the ICL resulted 0.23), while the $NN$-EV led to a quite fuzzy classification (the entropy term resulted 12.39). 
We checked also the AIC for both models, and this agreed with ICL.
Thus, $NN$-VE will be the only linear CWM considered hereafter. 
In the family of parsimonious mixtures of Gaussian distributions, the best model resulted EEE (homoscedastic group-covariance matrices; see \citealp{Fral:Raft:Murp:Scru:mclu:2012} for details).
Thus, we compared the performance of three Gaussian-based models whose classification results are reported in \tablename~\ref{tab:f.voles classification}.
%
\begin{table}[!ht] 
\centering
\begin{tabular}{cccc}
  \toprule
  & 
\begin{tabular}{ccc}
CWM
\\ $NN$-VE
\end{tabular}
& \texttt{flexmix} &
\begin{tabular}{ccc}
\texttt{mclust}\\
model EEE
\end{tabular}
 \\
 \midrule
ARI                  & 1.00000 & 0.02430 & 0.90810 \\
misclassification error  & 0.00000 & 0.40698 & 0.02326  \\
\bottomrule
\end{tabular}
\caption{f.voles data: 
classification results using different mixture-based approaches ($G=2$). 
}
\label{tab:f.voles classification}
\end{table}
The finite mixture of Gaussian regressions was the worse approach, reporting a  misclassification rate of 0.40698.
On the contrary, and surprisingly, our model $NN$-VE attains a perfect classification of the data
(we remark  the same optimal classification performance was obtained by all the ``-VE'' models in our family).

In conclusion, this is an example of ``strong'' assignment dependence where the group structure only depends by a different distribution of the covariates between the two groups (see also Proposition~\ref{lem:CWM vs. FMD}).

  
\section{Comparing the BIC and the ICL}
\label{sec:Comparing the BIC and the ICL}

A simulation study is described for comparing the performance of the BIC and the ICL with regard to the proposed family of models.
Five scenarios  are presented where data are simulated according to the following models: $NN$-EV, $NN$-VE, $NN$-VV, $Nt$-VE, and $tN$-EV.
In each scenario, 50 data sets of size $n=400$ are simulated with: $d=1$, $G=2$, and varying parameters.
The choice of considering different parameters is made to avoid particular configurations which may favor one of the competitive model selection criteria.

In each replication, the generating (true) model is specified as follows:
\begin{itemize}
	\item the mixture weight $\pi_1$ is randomly generated by a uniform distribution on $\left[0.2,0.8\right]$;
	\item as the variable $X$ is concerned, we refer to equation \eqref{eq:tCWM x}.
	Note that, we prefer to leave the matrix notation of the parameters $\bmu_g$ and $\bSigma_g$ even if, being $d=1$, they are indeed scalar values.
	In particular
\begin{itemize}
	\item if the model assumes $\bmu_1\neq \bmu_2$, then $\bmu_1$ and $\bmu_2$ are randomly generated by a standard normal distribution.
	If $\bmu_1=\bmu_2=\bmu$, then $\bmu$ is drawn by a standard normal distribution;
	\item if the model assumes $\bSigma_1\neq \bSigma_2$, then $\bSigma_1$ and $\bSigma_2$ are randomly generated by a $\chi^2_1$ distribution.
	If $\bSigma_1=\bSigma_2=\bSigma$, then $\bSigma$ is drawn by a $\chi^2_1$ distribution;
	\item if $p\left(\bx|\Omega_1\right)$ and $p\left(\bx|\Omega_2\right)$ are assumed to be $t$;
\begin{itemize}
	\item if the model assumes $\nu_1\neq\nu_2$, then $\nu_1$ and $\nu_2$ are randomly generated by a uniform distribution on $\left[2,5\right]$;
	\item if the model assumes $\nu_1=\nu_2=\nu$, then $\nu$ is drawn by a uniform distribution on $\left[2,5\right]$;
\end{itemize}
\end{itemize}
	\item as the variable $Y$ is concerned, by referring to equation \eqref{eq:tCWM y|x}, we have that
\begin{itemize}
	\item if the model assumes $\beta_{01}\neq \beta_{02}$ and $\beta_{11}\neq \beta_{12}$, then $\beta_{01}$ and $\beta_{02}$ are randomly generated by a standard normal distribution while $\beta_{11}$ and $\beta_{12}$ are drawn from a uniform distribution on $\left[-2,2\right]$.
	If $\beta_{01}=\beta_{02}=\beta_0$ and $\beta_{11}=\beta_{12}=\beta_1$, then $\beta_0$ is generated by a standard normal distribution and $\beta_1$ is drawn from a uniform distribution on $\left[-2,2\right]$;
	\item if the model assumes $\sigma^2_1\neq \sigma^2_2$, then $\sigma^2_1$ and $\sigma^2_2$ are randomly generated by a $\chi^2_1$ distribution.
	If $\sigma^2_1=\sigma^2_2=\sigma^2$, then $\sigma^2$ is generated by a $\chi^2_1$ distribution;
	\item if $p\left(y|\bx,\Omega_1\right)$ and $p\left(y|\bx,\Omega_2\right)$ are assumed to be $t$
\begin{itemize}
	\item if the model assumes $\zeta_1\neq\zeta_2$, then $\zeta_1$ and $\zeta_2$ are randomly generated by a uniform distribution on $\left[2,5\right]$;
	\item if the model assumes $\zeta_1=\zeta_2=\zeta$, then $\zeta$ is drawn by a uniform distribution on $\left[2,5\right]$.
\end{itemize}
\end{itemize}
\end{itemize}
The defined models guarantee various degrees of overlap between groups according to the generated parameters.

\begin{figure}[!ht]
\centering
\resizebox{0.76\textwidth}{!}{
\includegraphics{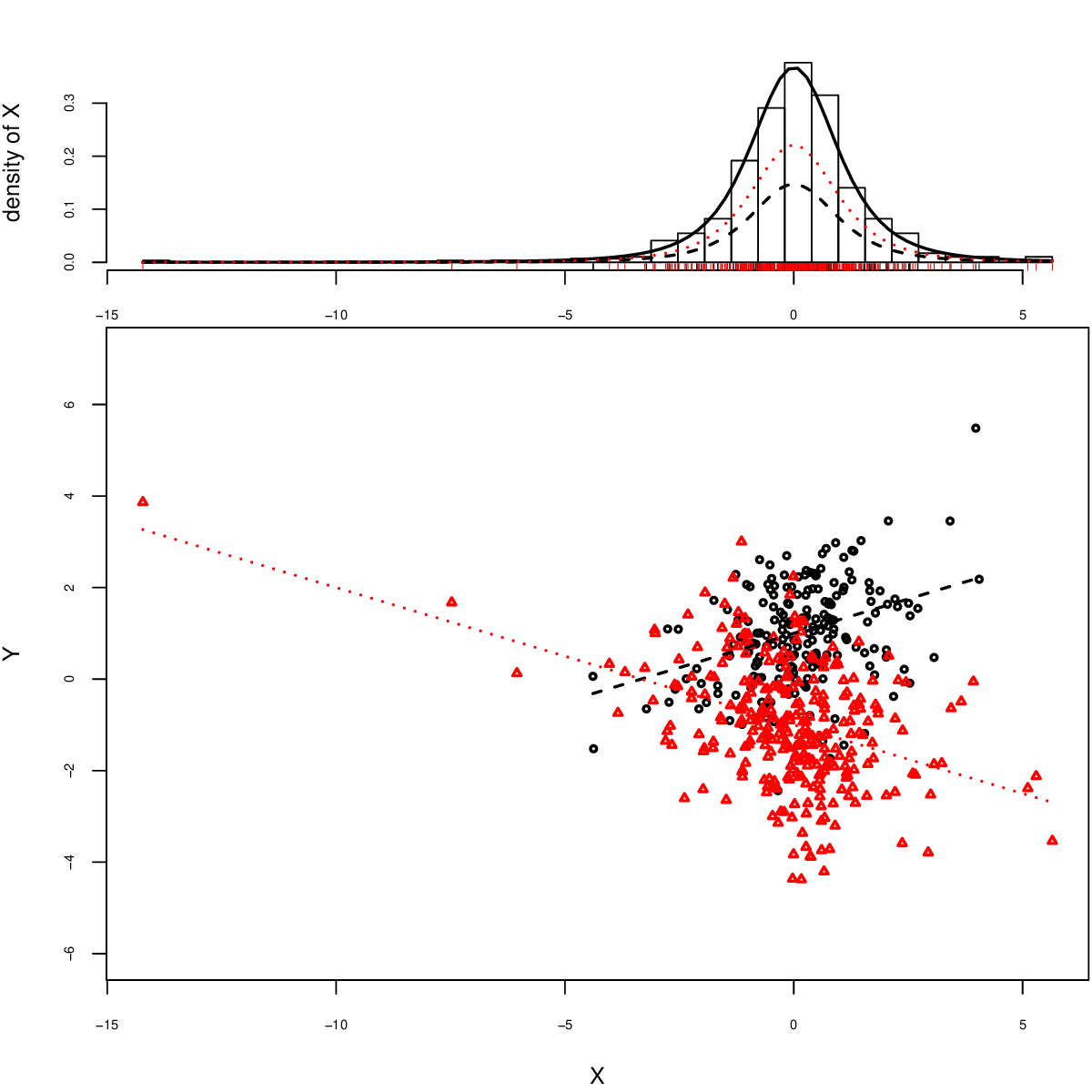} 
}
\caption{
CW-plot of data randomly generated from a $tN$-EV model with $G=2$.
}
\label{fig:BIC.tN-EV}
\end{figure}

\begin{figure}[!ht]
\centering
\resizebox{0.76\textwidth}{!}{
\includegraphics{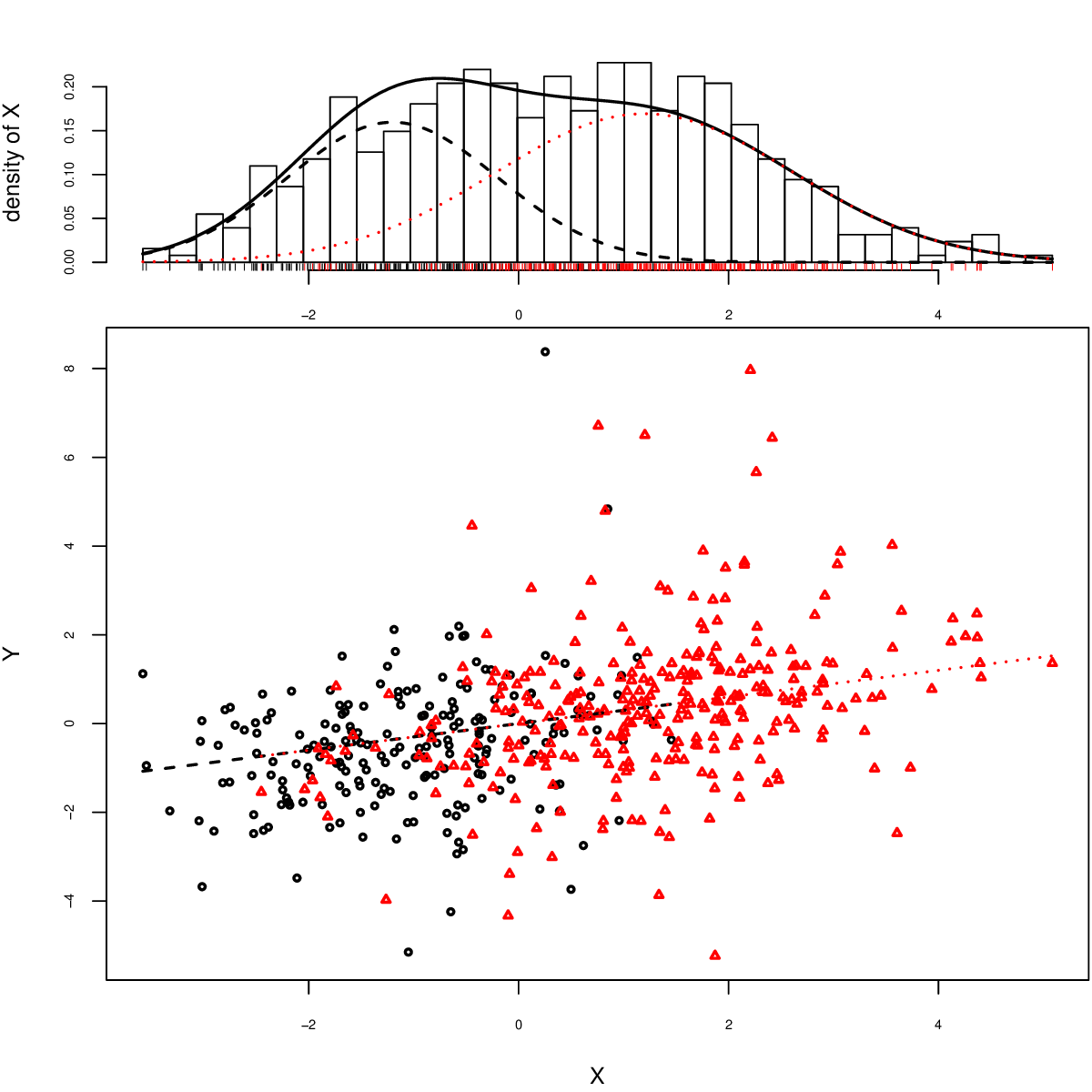} 
}
\caption{
CW-plot of data randomly generated from a $Nt$-VE model with $G=2$.
}
\label{fig:BIC.Nt-VE}
\end{figure}

In each replication, the true model is adopted to generate the data set; thus, all the 12 models are fitted with $G\in\left\{1,2,3\right\}$, leading to a total of 36 fitted models.  
\tablename~\ref{tab:SimBICresults} and \tablename~\ref{tab:SimICLresults} show the results for the BIC and the ICL, respectively.
Here, a value in position $\left(i,j\right)$ has to be read as ``number of times that the combination $\left(\text{model},\text{number of groups}\right)$ on column $j$ is selected to fit the true model (with $G=2$) on row $i$''.
Bold numbers highlight the number of times that the pair $\left(\text{true model},G=2\right)$ is selected.
\begin{sidewaystable}
\centering
\resizebox*{1\textwidth}{!}{
\begin{tabular}{c@{\hspace{-5mm}}rccccccccccccccccccccccccccccccccccc}
\toprule
	&	Fitted &	\multicolumn{3}{c}{$NN$-EV}			&		&	\multicolumn{3}{c}{$NN$-VE}			&		&	\multicolumn{3}{c}{$NN$-VV}			&		&	\multicolumn{3}{c}{$Nt$-EV}			&		&	\multicolumn{3}{c}{$Nt$-VE}			&		&	\multicolumn{3}{c}{$Nt$-VV}			&		&	\multicolumn{3}{c}{$tN$-EV}			&		&	\multicolumn{3}{c}{$tN$-VE}			&		&	\multicolumn{3}{c}{$tN$-VV}			\\													True	&	$G$ 	&	1	&2	&3	&		&	1	&2&	3	&		&	1	&2&	3	&		&	1	&2&	3	&		&	1	&2&	3	&		&	1&	2&	3	&		&	1	&2&	3	&		&	1	&2&	3	&		&	1	&2&	3	\\	\midrule
$NN$-EV	&		&	8	&\textbf{40}	&0	&		&	8	&0	&0	&		&	8	&0	&0	&		&	2	&0	&0	&		&	2	&0&	0	&		&	2	&0	&0	&		&	0&	0&	0	&		&	0	&0&	0	&		&	0	&0&	0	\\
$NN$-VE	&		&	5	&0	&0	&		&	5	&\textbf{40}	&0	&		&	5	&0	&0	&		&	4	&0	&0	&		&	4	&0&	0	&		&	4	&0	&0	&		&	0	&0&	0	&		&	0	&1&	0	&		&	0	&0	&0	\\
$NN$-VV	&		&	0	&0	&0	&		&	0	&0	&0	&		&	0	&\textbf{50}	&0	&		&	0	&0	&0	&		&	0	&0&	0	&		&	0	&0	&0	&		&	0&	0&	0	&		&	0	&0&	0	&		&	0	&0	&0	\\
$Nt$-VE	&		&	0	&1	&0	&		&	0	&0	&0	&		&	0	&0	&0	&		&	5	&0	&0	&		&	5	&\textbf{43}&	1	&		&	5	&0 &	0	&		&	0&	0&	0	&		&	0	&0&	0	&		&	0	&0&	0	\\
$tN$-EV	&		&	0	&0	&0	&		&	0	&1	&0	&		&	0	&0	&0	&		&	0	&0	&0	&		&	0	&0&	0	&		&	0	&0	&0	&		&	2	&\textbf{47}&	0	&		&	2	&0&	0	&		&	2	&0&	0	\\
\bottomrule
\end{tabular}
}
\caption{
Simulation results for the BIC.
Values in the table show the number of times, over 50 replications, that the model, and number of groups, on the column are selected to fit the true model (with two groups) which appears in the corresponding row.
Bold numbers highlight the largest number of times that the model selection criteria selects the true model.
} 
\label{tab:SimBICresults}
\end{sidewaystable}
\begin{sidewaystable}
\centering
\resizebox*{1\textwidth}{!}{
\begin{tabular}{c@{\hspace{-5mm}}rccccccccccccccccccccccccccccccccccc}
\toprule
	&	Fitted &	\multicolumn{3}{c}{$NN$-EV}			&		&	\multicolumn{3}{c}{$NN$-VE}			&		&	\multicolumn{3}{c}{$NN$-VV}			&		&	\multicolumn{3}{c}{$Nt$-EV}			&		&	\multicolumn{3}{c}{$Nt$-VE}			&		&	\multicolumn{3}{c}{$Nt$-VV}			&		&	\multicolumn{3}{c}{$tN$-EV}			&		&	\multicolumn{3}{c}{$tN$-VE}			&		&	\multicolumn{3}{c}{$tN$-VV}			\\													True	&	$G$ 	&	1	&2	&3	&		&	1	&2&	3	&		&	1	&2&	3	&		&	1	&2&	3	&		&	1	&2&	3	&		&	1&	2&	3	&		&	1	&2&	3	&		&	1	&2&	3	&		&	1	&2&	3	\\	\midrule
$NN$-EV	&		&	14	&\textbf{27}	&0	&		&	14	&0&	0	&		&	14&	0&	0	&		&	9	&0&	0	&		&	9	&0&	0	&		&	9	&0&	0	&		&	0	&0&	0	&		&	0	&0&	0	&		&	0	&0&	0	\\
$NN$-VE	&		&	13	&0	&0	&		&	13	&\textbf{24}&	0	&		&	13	&0&	0	&		&	0	&0&	0	&		&	0	&0&	0	&		&	0	&0&	0	&		&	12	&0&	0	&		&	12	&1&	0	&		&	12&	0&	0	\\
$NN$-VV	&		&	0	&0&	0	&		&	0	&0&	0	&		&	0	&\textbf{47}&	0	&		&	1	&0&	0	&		&	1	&0&	0	&		&	1	&0&	0	&		&	2	&0&	0	&		&	2	&0&	0	&		&	2	&0&	0	\\
$Nt$-VE	&		&	0	&0&	0	&		&	0	&0&	0	&		&	0	&0&	0	&		&	15	&0&	0	&		&	15	&\textbf{30}&	1	&		&	15	&0&	0	&		&	2	&2&	0	&		&	2	&0&	0	&		&	2	&0&	0	\\
$tN$-EV	&		&	0	&0&	0	&		&	0	&0&	0	&		&	0	&1&	0	&		&	0	&0&	0	&		&	0	&0&	0	&		&	0	&0&	0	&		&	16	&\textbf{33}&	0	&		&	16	&0&	0	&		&	16	&0&	0	\\
\bottomrule
\end{tabular}
}
\caption{
Simulation results for the ICL.
Values in the table show the number of times, over 50 replications, that the model, and number of groups, on the column are selected to fit the true model (with two groups) which appears in the corresponding row.
Bold numbers highlight the number of times that the model selection criteria selects the true model.
} 
\label{tab:SimICLresults}
\end{sidewaystable}
Note that: the columns referred to models of type ``$tt$-'' are missing simply because they have never been selected, and
the sum by row is greater than 50 because, when $G=1$ is selected, there is not difference between ``-VV'', ``-VE'', and ``-EV'' (see Section~\ref{sec:The parsimonious Student-t LCWMs}).
By comparing the results in these tables, the BIC seems to perform better than the ICL.
In particular, the ICL selects models with only one group a larger number of times than the BIC. 
This is probably induced by the scheme of definition of the true model that allows for groups with a strong overlap; thus, the entropy term of the ICL carries out a strong penalization which leads to the choice $G=1$.
\figurename~\ref{fig:BIC.tN-EV} and \figurename~\ref{fig:BIC.Nt-VE} display two examples where this happens.
From these examples we understand as it is difficult to establish the best model selection criterion; indeed, the ICL may be seen as better if the user actually does not want to separate two mixture components that are so similar that they do not constitute two different clusters in terms of interpretation. 
So, in general, it depends on the meaning of the data which criterion is better.



\section{Conclusions and discussion}
\label{sec:conclusions}

In this paper, a novel family of twelve linear cluster-weighted models was presented. 
Such a family represents a flexible and powerful tool for model-based clustering.
Maximum likelihood parameter estimation was performed according to the EM algorithm and model selection was accomplished using both the BIC and ICL.
Many computational aspects were illustrated and a simple, but very effective, hierarchical random initialization method was introduced. 
Model-based clustering, using the proposed family, was appreciated on the grounds of some applications to real data. 
Here, it is interesting to note how the data set related to the survey of students in Section~\ref{subsec:Student Data} justifies and motivates the search for a model in the proposed family.

Future work will involve the extension of the proposed family to the model-based classification context.
Moreover, the identifiability issue needs to be adequately addressed; a reference point is given by \citet{Henn:Iden:2000}.
Finally, Section~\ref{sec:Comparing the BIC and the ICL} presented first results to find out a suitable model selection criterion and motivates further research in this direction.


\section*{Acknowledgements}
The authors sincerely thank the Associate Editor and the referees for very helpful comments and valuable suggestions that have contributed to improving the quality of the manuscript.

\appendix

\section[EM-constraints for parsimonious models]{EM-constraints for parsimonious models}
\label{app:Constraints for parsimonious models}

In the following we describe how to impose constraints on the EM algorithm, described in Section~\ref{sec:Estimation via the EM algorithm} for the most general model $tt$-VV, to obtain parameter estimates for all the other models in \tablename~\ref{tab:parsimonious models}.
To this end, the itemization given at the beginning of Section~\ref{sec:The parsimonious Student-t LCWMs} will be considered as a benchmark scheme.

\subsection{Common $t$ for the component marginal densities}
\label{app:Common t for the component marginal densities}

When we constrain all the groups to have a common $t$ distribution for $\bX$, we have $\bmu_1=\cdots=\bmu_G=\bmu$, $\bSigma_1=\cdots=\bSigma_G=\bSigma$, and $\nu_1=\cdots=\nu_G=\nu$.
Thus, in the $\left(k+1\right)$th iteration of the EM algorithm, equations \eqref{eq:EU} and \eqref{eq:ElnU} must be replaced by
\begin{equation}
u_n^{\left(k\right)}=\frac{\nu^{\left(k\right)}+d}{\nu^{\left(k\right)}+\delta\left(\bx_n,\bmu^{\left(k\right)};\bSigma^{\left(k\right)}\right)}
\label{eq:EU constrained}
\end{equation}
and 
\begin{displaymath}
\widetilde{u}_n^{\left(k\right)} = \ln u_n^{\left(k\right)} + \psi\left(\frac{\nu^{\left(k\right)} + d}{2}\right) - \ln \left(\frac{\nu^{\left(k\right)} + d}{2}\right),
\end{displaymath}
respectively.
Furthermore, noting that $\sum_g\tau_{ng}=1$, equations \eqref{eq:Q4} and \eqref{eq:Q5} can be rewritten as 
\begin{equation}
Q_4\left(\boldsymbol{\vartheta};\text{\textsubtilde{$\boldsymbol{\psi}$}}^{\left(k\right)}\right)=\sum_{n=1}^NQ_{4n}\left(\boldsymbol{\vartheta};\text{\textsubtilde{$\boldsymbol{\psi}$}}^{\left(k\right)}\right)
\label{eq:Q4 constrained}
\end{equation}
and
\begin{displaymath}
Q_5\left(\nu;\text{\textsubtilde{$\boldsymbol{\psi}$}}^{\left(k\right)}\right)=\sum_{n=1}^NQ_{5n}\left(\nu;\text{\textsubtilde{$\boldsymbol{\psi}$}}^{\left(k\right)}\right),
\end{displaymath}
respectively, where
\begin{displaymath}
Q_{4n}\left(\boldsymbol{\vartheta};\text{\textsubtilde{$\boldsymbol{\psi}$}}^{\left(k\right)}\right)=\frac{1}{2}\left[-d\ln\left(2\pi\right)+d\widetilde{u}_n^{\left(k\right)}-\ln\left|\bSigma\right|-u_n\delta\left(\bx_n,\bmu;\bSigma\right)\right]
\end{displaymath}
and
\begin{displaymath}
Q_{5n}\left(\nu;\text{\textsubtilde{$\boldsymbol{\psi}$}}^{\left(k\right)}\right)=-\ln\Gamma\left(\frac{\nu}{2}\right)+\frac{\nu}{2}\ln \frac{\nu}{2}+\frac{\nu}{2}\left[\widetilde{u}_n^{\left(k\right)}-\ln u_n^{\left(k\right)}+\sum_{n=1}^N\left(\ln u_n^{\left(k\right)}-u_n^{\left(k\right)} \right)\right].
\end{displaymath}
Maximization of \eqref{eq:Q4 constrained}, with respect to $\boldsymbol{\vartheta}$, leads to
\begin{displaymath}
\bmu^{\left(k+1\right)} = \displaystyle\sum_{n=1}^Nu_n^{\left(k\right)}\bx_n\Big/\displaystyle\sum_{n=1}^Nu_n^{\left(k\right)}\end{displaymath}
and
\begin{displaymath}
\bSigma^{\left(k+1\right)} = \displaystyle\sum_{n=1}^Nu_n^{\left(k\right)}\left(\bx_n-\bmu^{\left(k+1\right)}\right)\left(\bx_n-\bmu^{\left(k+1\right)}\right)'\Big/\displaystyle\sum_{n=1}^Nu_n^{\left(k\right)}.
\end{displaymath}
For the updating of $\nu$, we need to numerically solve the equation
\begin{displaymath}
\sum_{n=1}^N\frac{\partial}{\partial \nu}Q_{5n}\left(\nu;\text{\textsubtilde{$\boldsymbol{\psi}$}}^{\left(k\right)}\right)=0,
\end{displaymath}
which corresponds to finding $\nu^{\left(k+1\right)}$ as the solution of
\begin{equation}
-\psi\left(\displaystyle\frac{\nu}{2}\right) + \ln\displaystyle\frac{\nu}{2}+ 1 + \displaystyle\displaystyle\sum_{n=1}^N\left(\ln u_n^{\left(k\right)}- u_n^{\left(k\right)} \right) +\psi\left(\displaystyle\frac{\nu^{\left(k\right)}+d}{2}\right)-\ln\left(\displaystyle\frac{\nu^{\left(k\right)}+d}{2}\right)=0.
\label{eq:nu constrained}
\end{equation}

\subsection{Common $t$ for the component conditional densities}
\label{app:Common t for the component conditional densities}

Similarly, when we constrain all the groups to have a common $t$ distribution for $Y|\bx$, we have $\bbeta_{11}=\cdots=\bbeta_{1G}=\bbeta_1$, $\beta_{01}=\cdots=\beta_{0G}=\beta_0$, $\sigma^2_1=\cdots=\sigma^2_G=\sigma^2$, and $\zeta_1=\cdots=\zeta_G=\zeta$.
Thus, in the $\left(k+1\right)$th iteration of the EM algorithm, equations \eqref{eq:EV} and \eqref{eq:ElnV} must be replaced by 
\begin{equation}
v_n^{\left(k\right)}=\frac{ \zeta^{\left(k\right)}+1 }{\zeta_g^{\left(k\right)} + \delta\left[y_n,\mu\left(\bx_n;\bbeta^{\left(k\right)}\right);\sigma^{2\left(r\right)}\right]
}
\label{eq:EV constrained}
\end{equation}
and
\begin{displaymath}
\widetilde{v}_n^{\left(k\right)}=\ln v_n^{\left(k\right)} + \psi\left(\frac{\zeta^{\left(k\right)} + 1}{2}\right) - \ln \left(\frac{\zeta^{\left(k\right)} + 1}{2}\right),
\end{displaymath}
respectively.
Also, equations \eqref{eq:Q2} and \eqref{eq:Q3} can be rewritten as
\begin{equation}
Q_2\left(\boldsymbol{\xi};\text{\textsubtilde{$\boldsymbol{\psi}$}}^{\left(k\right)}\right)=\sum_{n=1}^NQ_{2n}\left(\boldsymbol{\xi};\text{\textsubtilde{$\boldsymbol{\psi}$}}^{\left(k\right)}\right) 
\label{eq:Q2 constrained}
\end{equation}
and 
\begin{displaymath}
Q_3\left(\zeta;\text{\textsubtilde{$\boldsymbol{\psi}$}}^{\left(k\right)}\right)=\sum_{n=1}^NQ_{3n}\left(\zeta;\text{\textsubtilde{$\boldsymbol{\psi}$}}^{\left(k\right)}\right),
\end{displaymath}
respectively, where
\begin{displaymath}
Q_{2n}\left(\boldsymbol{\xi};\text{\textsubtilde{$\boldsymbol{\psi}$}}^{\left(k\right)}\right)=\frac{1}{2}\left\{-\ln\left(2\pi\right)+\widetilde{v}_n^{\left(k\right)}-\ln\sigma^2-v_n\delta\left[y_n,\mu\left(\bx_n;\bbeta\right);\sigma^{2}\right]\right\}
\end{displaymath}
and
\begin{displaymath}
Q_{3n}\left(\zeta;\text{\textsubtilde{$\boldsymbol{\psi}$}}^{\left(k\right)}\right)=-\ln\Gamma\left(\frac{\zeta}{2}\right)+\frac{\zeta}{2}\ln \frac{\zeta}{2}+\frac{\zeta}{2}\left[\widetilde{v}_n^{\left(k\right)}-\ln v_n^{\left(k\right)}+\sum_{n=1}^N\left(\ln v_n^{\left(k\right)}-v_n^{\left(k\right)} \right)\right].
\end{displaymath}
Maximization of \eqref{eq:Q2 constrained}, with respect to $\boldsymbol{\xi}$, leads to the updates
\begin{eqnarray*}
\bbeta_{1}^{\left(k+1\right)}&=&\left(\frac{\displaystyle\sum_{n=1}^Nv_n^{\left(k\right)}\bx_n\bx_n'}{\displaystyle\sum_{n=1}^Nv_n^{\left(k\right)}}-\frac{\displaystyle\sum_{n=1}^Nv_n^{\left(k\right)}\bx_n}{\displaystyle\sum_{n=1}^Nv_n^{\left(k\right)}
}\frac{\displaystyle\sum_{n=1}^Nv_n^{\left(k\right)}\bx_n'}{\displaystyle\sum_{n=1}^Nv_n^{\left(k\right)}
}\right)^{-1}
\cdot\nonumber\\
&&\cdot\left(
\frac{\displaystyle\sum_{n=1}^Nv_n^{\left(k\right)}y_n\bx_n}{\displaystyle\sum_{n=1}^Nv_n^{\left(k\right)}}
-
\frac{\displaystyle\sum_{n=1}^Nv_n^{\left(k\right)}y_n}{\displaystyle\sum_{n=1}^Nv_n^{\left(k\right)}
}\frac{\displaystyle\sum_{n=1}^Nv_n^{\left(k\right)}\bx_n}{\displaystyle\sum_{n=1}^Nv_n^{\left(k\right)}
}\right),
\end{eqnarray*}
\begin{displaymath}
\beta_0^{\left(k+1\right)}=\displaystyle \frac{\displaystyle\sum_{n=1}^Nv_n^{\left(k\right)}y_n}{
\displaystyle\sum_{n=1}^Nv_n^{\left(k\right)}}
-\bbeta_1^{\left(k+1\right)'}
\frac{\displaystyle\sum_{n=1}^Nv_n^{\left(k\right)}\bx_n}{\displaystyle\sum_{n=1}^Nv_n^{\left(k\right)}
}
\end{displaymath}
and
\begin{displaymath}
\sigma^{2\left(k+1\right)}=
\displaystyle\displaystyle\sum_{n=1}^Nv_n^{\left(k\right)}\left[y_n-\left(\beta_0^{\left(k+1\right)}+\boldsymbol{\beta}_1^{\left(k+1\right)'}\bx_n\right)\right]^2\Big/\displaystyle\sum_{n=1}^Nv_n^{\left(k\right)}.
\end{displaymath}
For the updating of $\zeta$, we need to numerically solve the equation
\begin{displaymath}
\sum_{n=1}^N\frac{\partial}{\partial \nu}Q_{3n}\left(\zeta;\text{\textsubtilde{$\boldsymbol{\psi}$}}^{\left(k\right)}\right)=0,
\end{displaymath}
which corresponds to finding $\zeta^{\left(k+1\right)}$ as the solution of
\begin{displaymath}
-\psi\left(\displaystyle\frac{\zeta}{2}\right) + \ln\displaystyle\frac{\zeta}{2}+ 1 + \displaystyle\displaystyle\sum_{n=1}^N\left(\ln v_n^{\left(k\right)}- v_n^{\left(k\right)} \right) +\psi\left(\displaystyle\frac{\zeta^{\left(k\right)}+1}{2}\right)-\ln\left(\displaystyle\frac{\zeta^{\left(k\right)}+1}{2}\right)=0.
\end{displaymath}

\subsection{Normal component marginal densities}
\label{app:Normal component marginal densities}

The normal case for the component distributions of $\bX$ can be obtained, as stated previously, as a limiting case when $\nu_g\rightarrow \infty$, $g=1,\ldots,G$.
Then, in \eqref{eq:EU}, $u_{ng}^{\left(k\right)}\rightarrow 1$.
Substituting this value into \eqref{eq:updated mean vector for X} and \eqref{eq:updated covariance matrix for X}, we obtain 
\begin{displaymath}
\bmu_g^{\left(k+1\right)} = \displaystyle\sum_{n=1}^N\tau_{ng}^{\left(k\right)}\bx_n\Big/\displaystyle\sum_{n=1}^N\tau_{ng}^{\left(k\right)}
\end{displaymath}
and
\begin{displaymath}
\bSigma_g^{\left(k+1\right)} = \displaystyle\sum_{n=1}^N\tau_{ng}^{\left(k\right)}\left(\bx_n-\bmu_g^{\left(k+1\right)}\right)\left(\bx_n-\bmu_g^{\left(k+1\right)}\right)'\Big/\displaystyle\sum_{n=1}^N\tau_{ng}^{\left(k\right)}.
\end{displaymath}
Naturally, in this case, we do not compute the additional $M$-step maximizing $Q_5\left(\text{\textsubtilde{$\nu$}};\text{\textsubtilde{$\boldsymbol{\psi}$}}^{\left(k\right)}\right)$ in \eqref{eq:Q5}.
Accordingly, for the sub-case $\bmu_1=\cdots=\bmu_G=\bmu$ and $\bSigma_1=\cdots=\bSigma_G=\bSigma$, in equation \eqref{eq:EU constrained} we have $u_n^{\left(k\right)}\rightarrow 1$ and the updated estimates of $\bmu$ and $\bSigma$ become      
\begin{displaymath}
\bmu = \displaystyle\frac{1}{n}\sum_{n=1}^N\bx_n 
\end{displaymath}
and
\begin{displaymath}
\bSigma = \displaystyle\frac{1}{n}\sum_{n=1}^N\left(\bx_n-\bmu\right)\left(\bx_n-\bmu\right)',
\end{displaymath}
which do not depend on the EM-iterations.

\subsection{Normal component conditional densities}
\label{app:Normal component conditional densities}

The normal case for the component distributions of $Y|\bX$ can be obtained as a limiting case when $\zeta_g\rightarrow \infty$, $g=1,\ldots,G$.
Then, in \eqref{eq:EV}, $v_{ng}^{\left(k\right)}\rightarrow 1$.
Substituting this value into \eqref{eq:updated beta1} and \eqref{eq:updated beta0}, we obtain
\begin{eqnarray*}
\bbeta_{1g}^{\left(k+1\right)}&=&\left(\frac{\displaystyle\sum_{n=1}^N\tau_{ng}^{\left(k\right)}\bx_n\bx_n'}{\displaystyle\sum_{n=1}^N\tau_{ng}^{\left(k\right)}}-\frac{\displaystyle\sum_{n=1}^N\tau_{ng}^{\left(k\right)}\bx_n}{\displaystyle\sum_{n=1}^N\tau_{ng}^{\left(k\right)}
}\frac{\displaystyle\sum_{n=1}^N\tau_{ng}^{\left(k\right)}\bx_n'}{\displaystyle\sum_{n=1}^N\tau_{ng}^{\left(k\right)}
}\right)^{-1}
\cdot\nonumber\\
&&\cdot\left(
\frac{\displaystyle\sum_{n=1}^N\tau_{ng}^{\left(k\right)}y_n\bx_n}{\displaystyle\sum_{n=1}^N\tau_{ng}^{\left(k\right)}}
-
\frac{\displaystyle\sum_{n=1}^N\tau_{ng}^{\left(k\right)}y_n}{\displaystyle\sum_{n=1}^N\tau_{ng}^{\left(k\right)}
}\frac{\displaystyle\sum_{n=1}^N\tau_{ng}^{\left(k\right)}\bx_n}{\displaystyle\sum_{n=1}^N\tau_{ng}^{\left(k\right)}
}\right),
\end{eqnarray*}
\begin{displaymath}
\beta_{0g}^{\left(k+1\right)} = \displaystyle \frac{\displaystyle\sum_{n=1}^N\tau_{ng}^{\left(k\right)}y_n}{
\displaystyle\sum_{n=1}^N\tau_{ng}^{\left(k\right)}}
-\bbeta_{1g}^{\left(k+1\right)'}
\frac{\displaystyle\sum_{n=1}^N\tau_{ng}^{\left(k\right)}\bx_n}{\displaystyle\sum_{n=1}^N\tau_{ng}^{\left(k\right)}
}	
\end{displaymath}
and
\begin{displaymath}
	\sigma^{2\left(k+1\right)}_g =
\displaystyle\displaystyle\sum_{n=1}^N\tau_{ng}^{\left(k\right)}\left[y_n-\left(\beta_{0g}^{\left(k+1\right)}+\boldsymbol{\beta}_{1g}^{\left(k+1\right)'}\bx_n\right)\right]^2\Big/\displaystyle\sum_{n=1}^N\tau_{ng}^{\left(k\right)}.
\end{displaymath}
We again do not compute the additional $M$-step maximizing $Q_3\left(\text{\textsubtilde{$\zeta$}};\text{\textsubtilde{$\boldsymbol{\psi}$}}^{\left(k\right)}\right)$ in \eqref{eq:Q3}.
Accordingly, for the sub-case $\bbeta_{11}=\cdots=\bbeta_{1G}=\bbeta_1$, $\beta_{01}=\cdots=\beta_{0G}=\beta_0$, and $\sigma^2_1=\cdots=\sigma^2_G=\sigma^2$, in equation \eqref{eq:EV constrained} we have $v_n^{\left(k\right)}\rightarrow 1$ and the updated estimates of $\bbeta_1$, $\beta_0$, and $\sigma^2$ become
\begin{displaymath}
\bbeta_1 =
\left(\displaystyle\frac{1}{n}\displaystyle\sum_{n=1}^N\bx_n\bx_n'-\frac{1}{n^2}\displaystyle\sum_{n=1}^N\bx_n\displaystyle\sum_{n=1}^N\bx_n'\right)^{-1}\left(
\displaystyle\frac{1}{n}\sum_{n=1}^Ny_n\bx_n
-\frac{1}{n^2}
\displaystyle\sum_{n=1}^Ny_n\displaystyle\sum_{n=1}^N\bx_n\right),
\end{displaymath}
\begin{displaymath}
\beta_0=\displaystyle \frac{1}{n}\displaystyle\sum_{n=1}^Ny_n-\frac{1}{n}\bbeta_1'
\displaystyle\sum_{n=1}^N\bx_n
\end{displaymath}  
and
\begin{displaymath}
	\sigma^2 =
\displaystyle\frac{1}{n}\displaystyle\sum_{n=1}^N\left[y_n-\left(\beta_0+\boldsymbol{\beta}_1'\bx_n\right)\right]^2,
\end{displaymath}
which do not depend on the EM-iterations.

\bibliographystyle{model5-names}
\bibliography{References}

\end{document}